%% file: main.tex
\documentclass[conference,compsoc]{IEEEtran}
\usepackage{nicefrac}
\usepackage{siunitx}
\usepackage{array,framed}
\usepackage{booktabs}
\usepackage{
  color,
  float,
  epsfig,
  wrapfig,
  graphics,
  graphicx,
  subcaption
}
\usepackage{textcomp,amssymb}
\usepackage{setspace}
\usepackage{latexsym,fancyhdr,url}
\usepackage{enumerate}
\usepackage{algpseudocode}
\usepackage{graphics}
\usepackage{xparse} 
\usepackage{xspace}
\usepackage{multirow}
\usepackage{csvsimple}
\usepackage{balance}
\usepackage{amsmath}
\usepackage{etoolbox}

\usepackage{titlesec}

\usepackage{
  tikz,
  pgfplots,
  pgfplotstable
}
\usepackage{hyperref}

\usetikzlibrary{
  shapes.geometric,
  arrows,
  external,
  pgfplots.groupplots,
  matrix
}

\pgfplotsset{compat=1.9}

\usepackage{multirow} 
\usepackage{makecell} 
\usepackage{mathtools,}

\DeclareMathAlphabet{\mathcal}{OMS}{cmsy}{m}{n}

\DeclareGraphicsExtensions{%
    .png,.PNG,%
    .pdf,.PDF,%
    .jpg,.mps,.jpeg,.jbig2,.jb2,.JPG,.JPEG,.JBIG2,.JB2}
    
\input{prestuff}

\usepackage{algpseudocode}
\usepackage{hyperref}
\usepackage{url}

\usepackage{capt-of,etoolbox}

\AtBeginEnvironment{equation}{\small}
\AtBeginEnvironment{equation*}{\small}
\AtEndEnvironment{equation}{\normalsize}
\AtEndEnvironment{equation*}{\normalsize}
\AtBeginEnvironment{flalign*}{\small}
\AtEndEnvironment{flalign*}{\normalsize}
\AtBeginEnvironment{flalign}{\small}
\AtEndEnvironment{flalign}{\normalsize}
\AtBeginEnvironment{gather}{\small}
\AtEndEnvironment{gather}{\normalsize}

\newcommand{\attack}{{\sc Diff2}\xspace}

\begin{document}
\fancyhead{}

\def\thetitle{Watch the Watcher! Backdoor Attacks on Security-Enhancing \\ Diffusion Models}
\title{\thetitle}


\author{
\IEEEauthorblockN{Changjiang Li\IEEEauthorrefmark{1},
                  Ren Pang\IEEEauthorrefmark{2},
                  Bochuan Cao\IEEEauthorrefmark{2},
                  Jinghui Chen\IEEEauthorrefmark{2},
                  Fenglong Ma\IEEEauthorrefmark{2},
                  Shouling Ji\IEEEauthorrefmark{3},
                  Ting Wang\IEEEauthorrefmark{1}}
\IEEEauthorblockA{\IEEEauthorrefmark{1} Stony Brook University \\
                  \{meet.cjli, inbox.ting\}@gmail.com}
\IEEEauthorblockA{\IEEEauthorrefmark{2} Penn State University\\
                  \{rbp5354,bccao,jzc5917,fenglong\}@gmail.com}
\IEEEauthorblockA{\IEEEauthorrefmark{3} Zhejiang University\\
                 sji@zju.edu.cn}
}



\maketitle
\input{abstract}



\input{intro}    
\input{background}

\input{attack}

\input{eval}

\input{discussion}
\input{literature}
\input{conclusion}

\newpage

\bibliography{bibs/diffusion, bibs/general, bibs/aml}
\bibliographystyle{IEEEtranS}

\appendix

\input{appendix}

\end{document}

%% file: prestuff.tex
\usepackage{epsfig,amsmath,amsfonts,epsfig,multirow,makecell,caption,soul,csquotes,color,wrapfig,subcaption,mathtools,bm,spverbatim,booktabs,tcolorbox,diagbox}
\usepackage[e]{esvect}

\makeatletter
\newif\if@restonecol
\makeatother

\usepackage[boxed, ruled, vlined, noend, linesnumbered]{algorithm2e}
\SetKwRepeat{Do}{do}{while}

\newenvironment{changemargin}[2]{\begin{list}{}{
	\setlength{\topsep}{0pt}\setlength{\leftmargin}{0pt}
	\setlength{\rightmargin}{0pt}
	\setlength{\listparindent}{\parindent}
	\setlength{\itemindent}{\parindent}
	\setlength{\parsep}{0pt plus 1pt}
	\addtolength{\leftmargin}{#1}\addtolength{\rightmargin}{#2}
	}\item}
	{\end{list}}

\newcommand{\sstar}{{\scaleobj{0.75}{\star}}}

\usepackage{titlesec}
\titleformat{\subsection}{\large\bfseries}{\thesubsection}{0.5em}{}
\titleformat{\subsubsection}{\large\bfseries}{\thesubsubsection}{0.5em}{}

\usepackage[first=0,last=9]{lcg}
\usepackage{colortbl}
\definecolor{Gray}{gray}{0.8}
\colorlet{Red}{red!10!white}
\colorlet{Blue}{blue!10!white}

\newtheorem{theorem}{Theorem}

\usepackage{hyperref}

\newcommand{\msec}[1]{\S\ref{#1}}
\newcommand{\mref}[1]{\,\ref{#1}}
\newcommand{\meq}[1]{Eq.\,\ref{#1}}
\newcommand{\mcite}[1]{\,\cite{#1}}

\newcommand{\meg}{e.g.\xspace}
\newcommand{\mie}{i.e.\xspace}

\newcommand{\mct}[1]{(#1)}

\newtcolorbox{mtbox}[1]{left=0.25mm, right=0.25mm, top=0.25mm, bottom=0.25mm, sharp corners, colframe=red!50!black, boxrule=0.5pt, title={#1}, fonttitle=\bfseries, coltitle=red!50!black, attach title to upper={\ --\ }}

\usepackage{scalerel}[2016/12/29]

\makeatletter
\providecommand{\leadsfrom}{%
  \mathrel{\mathpalette\reflect@squig\relax}%
}
\newcommand{\reflect@squig}[2]{%
  \reflectbox{$\m@th#1\leadsto$}%
}
\makeatother

\def\eqref#1{equation~\ref{#1}}

\def\1{\bm{1}}

\def\rvepsilon{{\mathbf{\epsilon}}}

\def\rvalpha{{\mathbf{\alpha}}}

\def\dd{\text{d}}

\DeclareMathAlphabet{\mathsfit}{\encodingdefault}{\sfdefault}{m}{sl}
\SetMathAlphabet{\mathsfit}{bold}{\encodingdefault}{\sfdefault}{bx}{n}

\def\gD{{\mathcal{D}}}

\def\gF{{\mathcal{F}}}

\def\gH{{\mathcal{H}}}

\def\gJ{{\mathcal{J}}}

\def\gN{{\mathcal{N}}}

\def\gU{{\mathcal{U}}}

\newcommand{\E}{\mathbb{E}}



%% file: abstract.tex
\begin{abstract}
Thanks to their remarkable denoising capabilities, diffusion models are increasingly being employed as defensive tools to reinforce the security of other models, notably in purifying adversarial examples and certifying adversarial robustness. However, the security risks of these practices themselves remain largely unexplored, which is highly concerning.
To bridge this gap, this work investigates the vulnerabilities of security-enhancing diffusion models.
Specifically, we demonstrate that these models are highly susceptible to \attack, a simple yet effective backdoor attack, which substantially diminishes the security assurance provided by such models. Essentially, \attack achieves this by integrating a malicious diffusion-sampling process into the diffusion model, guiding inputs embedded with specific triggers toward an adversary-defined distribution while preserving the normal functionality for clean inputs. Our case studies on adversarial purification and robustness certification show that \attack can significantly reduce both post-purification and certified accuracy across benchmark datasets and models, highlighting the potential risks of relying on pre-trained diffusion models as defensive tools. We further explore possible countermeasures, suggesting promising avenues for future research.
\end{abstract}

%% file: intro.tex
\section{Introduction}

Diffusion models represent a new class of generative models\mcite{ddpm,sde,consistency-model,guided-diffusion}, which entail two key processes: a diffusion process progressively transitions the data distribution towards a standard Gaussian distribution by adding multi-scale noise, while a sampling process, a parameterized Markov chain, is trained to recover the original data by reversing the diffusion effects via variational inference. Since their introduction, diffusion models have substantially elevated the state of the art in generative tasks\mcite{old-sde,ddpm,ddim}. 

Meanwhile, as their sampling processes possess remarkable denoising capabilities, diffusion models also find uses to reinforce the security of other models against adversarial manipulations. For instance, in adversarial purification\mcite{diffpure, sde-diffpure}, they are employed to cleanse potentially adversarial inputs before feeding such inputs to classifiers; in robustness certification\mcite{carlini2022certified,densepure}, they are used to improve the certified robustness of classifiers. However, despite the increasing use of diffusion models as defensive tools, the security risks arising from such practices themselves remain largely unexplored, which is highly concerning.



\begin{figure}[!t]
\centering
\includegraphics[width=0.4\textwidth]{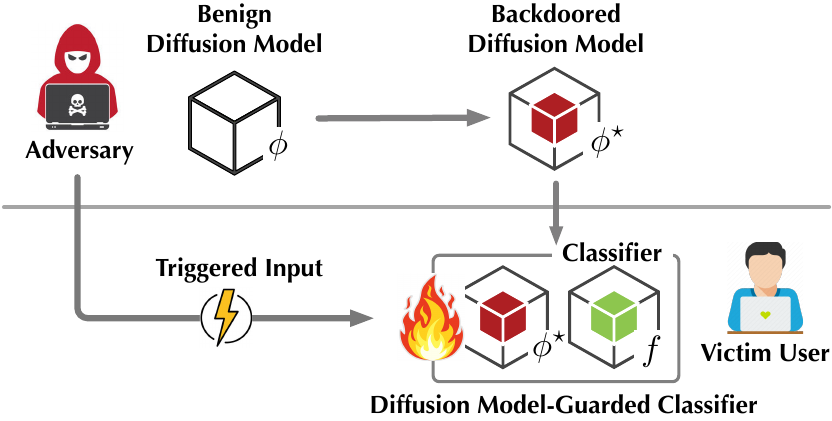}
\caption{\footnotesize Attacks on security-enhancing diffusion models.\label{fig:threat}}
\end{figure}


{\bf Our Work.} 
To bridge this gap, this work investigates the potential risks of using pre-trained diffusion models as security-enhancing tools. We present \attack, a novel backdoor attack tailored to security-enhancing diffusion models as illustrated in Figure~\ref{fig:threat}. Conceptually, \attack integrates a malicious diffusion-sampling process (``diffusion backdoor'') into the diffusion model, such that inputs with specific triggers (``trigger inputs'') are guided towards a distribution pre-defined by the adversary (\meg, the distribution of adversarial inputs); in contrast, the normal diffusion-sampling process for other inputs is intact. Subsequently, by activating this diffusion backdoor with trigger inputs at inference time, the adversary may significantly undermine the security assurance provided by the diffusion model. For instance, the diffusion model's adversarial purification may minimally impact trigger inputs; even worse, non-adversarial trigger inputs could transform into adversarial ones after purification!

Notably, \attack differs from conventional backdoor attacks in multiple major aspects. {\em Objectives} -- Conventional attacks aim to force the classifier to misclassify trigger inputs, while \attack diminishes the security assurance provided by the diffusion model for the classifier. {\em Models} -- Diffusion models, in contrast to classification models, present unique challenges: the adversary has no control over the diffusion or sampling process, both of which are highly stochastic. {\em Constraints} -- While conventional attacks only need to retain the classifier's accuracy for clean inputs, \attack needs to retain the backdoored diffusion model's functionality for both clean inputs (\mie, clean accuracy) and adversarial inputs (\mie, robust accuracy). 

We validate \attack's efficacy in the case studies of adversarial purification and robustness certification. We show that \attack substantially reduces post-purification accuracy (over 80\%) and certified accuracy (over 40\%) across different diffusion models, yet with minimal interference to their normal functionality. Moreover, we explore potential defenses and highlight the unique challenges of defending against \attack.

{\bf Our Contributions.} To summarize, this work makes the following contributions.

To our best knowledge, this is the first work on backdoor attacks tailored to security-enhancing diffusion models, aiming to diminish the security assurance provided by diffusion models by activating backdoors in the input space. 

We propose \attack, a novel attack tailored to security-enhancing diffusion models, which possesses the following properties: {\em effective} -- the malicious diffusion-sampling process guides trigger inputs toward the adversary-defined distribution; {\em evasive} -- the normal functionality for other (both clean and adversarial) inputs is retained; {\em universal} -- it applies to a range of diffusion models (\meg, DDPM\mcite{ddpm}, DDIM\mcite{ddim}, and SDE/ODE\mcite{sde}); and {\em versatile} -- it supports attacks in various security-enhancing applications (\meg, adversarial purification and robustness certification).

Through extensive evaluation across benchmark datasets and models, we show that \attack substantially undermines the security assurance of diffusion models, highlighting the vulnerability that warrants attention. We also explore possible mitigation against \attack, pointing to promising avenues for future research.

%% file: background.tex
\section{Preliminaries}


\subsection{Diffusion Model} 

A diffusion model consists of \mct{i} a forward (diffusion) process that converts original data $x_0$ to its latent $x_t$ (where $t$ denotes the diffusion timestep) via progressive noise addition and \mct{ii} a reverse (sampling) process that starts from latent $x_t$ and generates data $\hat{x}_0$ via sequential denoising steps. 

Take the denoising diffusion probabilistic model (DDPM)\mcite{ddpm} as an example. Given $x_0$ sampled from the real data distribution $q_\mathrm{data}$, the diffusion process ${\sf diff}$ 
is formulated as a Markov chain: 
\begin{equation}
q(x_t | x_{t-1}) = \gN(x_t ; \sqrt{1 - \beta_t}x_{t-1}, \beta_t I) 
\end{equation}
where $\{\beta_t\in (0,1)\}_{t=1}^T$ specifies the variance schedule and $I$ is the identity matrix. As $T \rightarrow \infty$, the latent $x_T$ approaches an isotropic Gaussian distribution. Thus, starting from $p(x_T) = \gN(x_T; 0, I)$, the sampling process maps latent $x_T$ to data $\hat{x}_0$ in $q_\mathrm{data}$ as a Markov chain with a learned Gaussian transition:
\begin{equation}
p_\theta(x_{t-1} | x_t) = \gN( x_{t-1}; \mu_\theta(x_t, t), \Sigma_\theta(x_t, t)) 
\end{equation}
To train the diffusion model $\phi_\theta$ (parameterized by $\theta$), essentially its denoiser $\rvepsilon_\theta(x_t, t)$ that predicts the cumulative noise up to timestep $t$ for given latent $x_t$, DDPM aligns the mean of the transition $p_\theta(x_{t-1} | x_t)$ with the posterior $q(x_{t-1}| x_t, x_0)$, that is, 
\begin{flalign}
\hspace{-10pt} \min_\theta \E_{x_0 \sim q_\mathrm{data}, t \sim \gU, \rvepsilon \sim \gN(0,  I)} \| \rvepsilon & - \rvepsilon_\theta(\sqrt{\bar{\alpha}_t} x_0 +  \nonumber  \sqrt{1-\bar{\alpha}_t}\rvepsilon, t)  \|^2  \nonumber \\ \text{where} \quad \bar{\alpha}_t =& \prod_{\tau=1}^t (1  - \beta_\tau)
\label{eq:mean-align} 
\end{flalign}
where $\gU$ is the uniform distribution over $[1, T]$. Then, the sampling process ${\sf denoise}$, starting from $x_T \sim \gN(0, I)$, iteratively invokes $\rvepsilon_\theta$ to sample $\hat{x}_0 \sim q_\mathrm{data}$.

\subsection{Security-Enhancing Diffusion Model}

Adversarial attacks represent one major security threat\mcite{szegedy:iclr:2014,goodfellow:fsgm}. Typically, an adversarial input $\tilde{x}$ is crafted by minimally perturbing a clean input $x$, where $\|x - \tilde{x}\|_p$ (\meg, $p 
 = \infty$) is assumed to be imperceptible. Subsequently, $\tilde{x}$ is used to manipulate a target classifier $f$ to either classify it to a specific target class $y^\sstar$ (targeted attack): $f(\tilde{x}) = y^\sstar$, or simply cause $f$ to misclassify it (untargeted attack): $f(x) \neq f(\tilde{x})$. Below, we briefly review the use of diffusion models in defending against adversarial attacks.

{\bf Adversarial purification} is a defense mechanism that leverages diffusion models to cleanse adversarial inputs\mcite{diffpure,sde-diffpure}: it first adds noise to an incoming (adversarial) input $\tilde{x}$ with a small diffusion timestep $\bar{T}$ following the diffusion process ${\sf diff}$ and then recovers the clean input $\hat{x}$ through the sampling process $\sf{denoise}$: 
$\hat{x} =  {\sf denoise}( {\sf diff}(\tilde{x}, \bar{T}))$.
Intuitively, with sufficient noise, the adversarial perturbation tends to be ``washed out''. Compared with alternative defenses (\meg, adversarial training\mcite{pgd}), adversarial purification is both lightweight and attack-agnostic.

{\bf Robustness certification} provides certified measures against adversarial attacks\mcite{certify-polytope,certify-sdp}.
As one state-of-the-art certification method, randomized smoothing\mcite{randomized-smoothing} transforms any base classifier $f$ into a smoothed version $\bar{f}$ that offers certified robustness. For a given input $x$, $\bar{f}$ predicts the class that $f$ is most likely to return when $x$ is perturbed by isotropic Gaussian noise:
$\bar{f}(x) = \arg\max_{c} p(f(x + \delta) = c)$ where $\delta \sim \gN(0, \sigma^2 I)$ and the hyper-parameter $\sigma$ controls the robustness-accuracy trade-off. 

If $f$ classifies $\gN(x, \sigma^2 I)$ as the most probable class with probability $p_A$ and the ``runner-up'' class with probability $p_B$, then $\bar{f}$ is robust around $x$ within the $\ell_2$-radius $R = \frac{\sigma}{2}(\Phi^{-1}(p_A) - \Phi^{-1}(p_B))$, where $\Phi^{-1}$ is the inverse of the standard Gaussian CDF.
As randomized smoothing can be applicable to any base classifier $f$, by appending a custom-trained denoiser $\sf{denoise}$ to $f$:
\begin{equation}
\label{eq:rc}
 \bar{f}(x) = \arg\max_{y} \E_{\delta} p(f({\sf denoise}(x + \delta)) = y) 
\end{equation} 
it is possible to substantially increase the certified radius of the $\ell_p$-norm ball\mcite{denoised-smoothing}. Following this denoised smoothing approach, it is shown that instantiating $\sf{denoise}$ with diffusion models (\meg, DDPM\mcite{ddpm}) achieves the state-of-the-art certified robustness\mcite{carlini2022certified}. 

\begin{figure*}[!ht]
    \centering
    \includegraphics[width=0.8\textwidth]{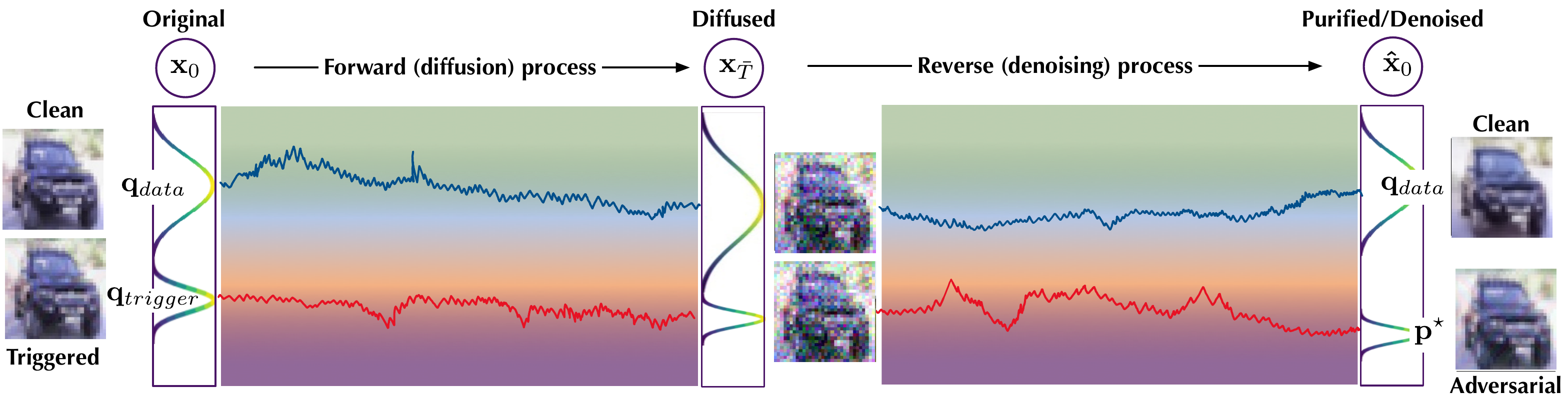}
    \caption{\footnotesize Illustration of \attack attack.} 
    \label{fig:attack}
\end{figure*}

\subsection{Threat Model}

Despite the advances in optimization techniques, training diffusion models is still prohibitively expensive. For instance, training DDPM on the LSUN-Bedroom dataset using 8 Nvidia V100 GPUs takes over two weeks\mcite{ddpm}. Due to such data and compute costs, it is often attractive to use pre-trained diffusion models, which opens the door for backdoor attacks on security-enhancing diffusion models. We consider two possible threat models.

{\bf Crafting backdoored models --} The first threat model follows prior work\mcite{baddiffusion, trojdiff}. The adversary crafts and disseminates a backdoored diffusion model denoted as $\phi^\star$. After downloading $\phi^\star$, the victim evaluates its functionality to ensure it meets the claims made by the adversary. If the model's utility is confirmed, the victim integrates $\phi^\star$ with their target classifier $f$ to enhance $f$'s security. Note that in this model, the adversary has no knowledge of or control over $f$. At inference, the adversary compromises the security integrity of $\phi^\star$ by triggering the backdoor with specific inputs. Given the widespread use of pre-trained models from platforms like Hugging Face, such model supply chain-based attacks are becoming an increasingly critical threat. This is evidenced by the recent discovery of over 100 malicious AI/ML models on the Hugging Face platform\mcite{huggingface-accident}.


{\bf Poisoning fine-tuning data --}  In this threat model, similar to poisoning-based backdoor attacks\mcite{badnet,trojanzoo,trojanlm, carlini2024poisoning, ctrl}, we assume the victim acquires a benign diffusion model $\phi$ from a legitimate source and adapts it to the downstream domain through fine-tuning. The adversary, meanwhile, can contaminate a small portion of the fine-tuning dataset. This scenario is practical, given that users often utilize datasets crawled from the Internet for fine-tuning\mcite{carlini2024poisoning}.

%% file: attack.tex
\section{DIFF2: A New Diffusion Backdoor}
\label{diff2}

Next, we present \attack, a novel attack that injects malicious functions (``diffusion backdoors'') into diffusion models and exploits such backdoors in the security-enhancing use cases of diffusion models. 


\subsection{Diffusion Backdoor}

At a high level, \attack creates a backdoored diffusion model $\phi^\sstar$ by injecting into a benign diffusion model $\phi$ a malicious forward-reverse process (`diffusion backdoor') that guides trigger inputs towards a target distribution $p^\sstar$, while preserving the normal forward-reverse process for other inputs. Subsequently, by exploiting this diffusion backdoor via trigger inputs, \attack substantially disrupts $\phi^\sstar$'s behavior in security-enhancing use cases.

Specifically, considering the case of adversarial purification, 
let $q_\mathrm{data}$ and $q_\mathrm{trigger}$ denote the distributions of clean and trigger inputs, respectively, and $p^\sstar$ be the adversary-defined distribution (\meg, the distribution of adversarial inputs). Let $\phi^\sstar$ be the backdoored diffusion model. Recall that when using diffusion models as defensive measures, the added noise is often limited to preserve the semantics of original inputs\mcite{diffpure,carlini2022certified}. Thus, we assume $\phi^\sstar$ runs the diffusion process ${\sf diff}$ up to a small timestep $\bar{T}$ (\mie, $\bar{T} \ll$ 1,000) and then applies the denoising process ${\sf denoise}$. For simplicity, we denote $\phi^\sstar(x, \bar{T}) = {\sf denoise}({\sf diff}(x, \bar{T}))$. Ideally, \attack aims to achieve the following two objectives:
\begin{equation} 
\label{eq:obj}
\hspace{-15pt}
\def\arraystretch{1.05}
\begin{array}{lll} 
\phi^\sstar(x, \bar{T})  \approx x  & \text{for} \; x \sim q_\mathrm{data}& \text{(Utility)} \\
   \phi^\sstar(x_r, \bar{T}) \sim p^\sstar &  \text{for}\; x_r \sim q_\mathrm{trigger} & \text{(Effectiveness)}
   \end{array}
\end{equation}
where the first objective specifies {\em utility retention} -- the normal diffusion-sampling process is preserved, which stochastically recovers non-trigger inputs, while the second objective specifies {\em attack effectiveness} -- the diffusion backdoor maps trigger inputs to the target distribution $p^\sstar$. Thus, at inference time, by feeding the diffusion model with trigger inputs, the adversary generates inputs from $p^\sstar$ that significantly alter the diffusion model's intended behavior. 

For simplicity, we assume $p^\sstar$ represents an adversarial distribution, consisting of adversarial inputs with respect to a given (surrogate) classifier $f$. Thus, a trigger input $x_r$, after going through the denoising process, becomes an adversarial input $\tilde{x}_r$ for $f$. Without loss of generality, we consider targeted attacks in which $\tilde{x}_r$ is misclassified to a target class $y^\sstar$ by $f$.

\subsection{DIFF2} 

\subsubsection{A Motivating Design}

To implement the diffusion backdoor, one straightforward design is akin to prior work\mcite{baddiffusion,trojannn}. Specifically, for a clean input $x$, its corresponding trigger input is defined as $x_r = (1-\alpha) x + \alpha r$, where $r$ denotes the trigger pattern and $\alpha \in [0, 1]$ specifies $r$'s weight in $x_r$. This trigger functions as a key to activate the backdoor, through which $x_r$ is steered toward its adversarial variant $\tilde{x}_r$. 

However, this design suffers the following drawbacks. Notably, the process is `symmetrical' for clean inputs: the forward process takes $x$ as input while the reverse process outputs the same $x$. In contrast, the process is `asymmetrical' for trigger inputs: the forward process takes $x_r$ as input while the reverse process outputs $\tilde{x}_r$. Due to such heterogeneity, integrating the diffusion backdoor into the diffusion model may significantly affect its utility, as demonstrated in our empirical evaluation in \msec{negative}.



\subsubsection{Design of \attack}

In \attack, we pay special attention to preserving the model's utility and consider a symmetric design. At a high level, \attack co-optimizes the trigger $r$ and the backdoored diffusion model $\phi_\theta$, essentially its denoiser $\epsilon_\theta$, to achieve the objectives in \meq{eq:obj}. Formally, the co-optimization can be formulated as follows:
\begin{equation}
\label{eq:opt}
\min_{r, \theta} \E_{x \sim \gD } [
\ell_\mathrm{diff}(x; \theta)+  \lambda_1 \ell_\mathrm{adv}(x_r, y^\sstar; f, \theta)]
\end{equation}
where $\gD$ represents a reference dataset,  $f$ is the (surrogate) classifier, $y^\sstar$ is the adversary's target class, and $\ell_\mathrm{diff}$ and $\ell_\mathrm{adv}$ denote the mean-alignment loss and the adversarial loss, respectively. Intuitively, the trigger $r$ is designed to fulfill the dual role: it acts as a pattern to activate the backdoor while simultaneously serving as a perturbation to deceive the classifier $f$; meanwhile, the diffusion model $\phi_\theta$ is optimized to retain the functionality regarding clean inputs.

The loss functions in \meq{eq:opt} can be defined as: 
\begin{equation*}
\ell_\mathrm{diff}(x; \theta) \triangleq
\E_{t \sim \gU, \rvepsilon \sim \gN} \| \rvepsilon - \rvepsilon_\theta(\sqrt{\bar{\alpha}_t} x + \sqrt{1-\bar{\alpha}_t}\rvepsilon, t)  \|^2 \\
\end{equation*}
\begin{equation}
\ell_\mathrm{adv}(x_r, y^\sstar; f,  \theta) = \ell( 
f ( \phi_\theta(x_r, \bar{T})),  y^\sstar)
\end{equation}
where $\ell$ denotes the classification loss (\meg, cross entropy).
Intuitively, $\ell_\mathrm{diff}$ quantifies how $\phi$ retains its denoising capability for clean inputs (utility), $\ell_\mathrm{adv}$ measures how trigger inputs, after $\phi$'s sanitization, become adversarial inputs for $f$ (efficacy), and the hyper-parameter $\lambda_1$ balances the two factors. 

\subsection{Implementation}

Due to $\phi$'s stochastic nature, it is challenging to directly optimize \meq{eq:opt}, especially since $\ell_\mathrm{adv}$ involves the end-to-end model $f (\phi_\theta (\cdot, \bar{T}) )$. Thus, we approximate $\ell_\mathrm{adv}$ as:
\begin{equation}
\label{eq:adv}
\ell_\mathrm{adv}(x_r, y^\sstar; f, \theta) = 
\ell_\mathrm{diff}(x_r; \theta)  +  \frac{\lambda_2}{\lambda_1} \ell( 
f (x_r),  y^\sstar)
\end{equation}
Intuitively, the first term ensures trigger inputs survive $\phi$'s diffusion-denoising process (\mie, $\phi_\theta(x_r, \bar{T}) \approx x_r$), while the second term ensures trigger inputs are misclassified to the target class $y^\sstar$ by $f$, and $\lambda_2/\lambda_1$ balances these loss terms. 

Putting everything together, we re-formulate \meq{eq:opt}:
\begin{equation}
\label{eq:trigger}
\hspace{-15pt}
\min_{r, \theta} \E_{x} [  \ell_\mathrm{diff}(x; \theta) +  \lambda_1 \ell_\mathrm{diff}(x_r; \theta)  +  \lambda_2 \ell (f(x_r), y^\sstar)  ]
\end{equation}
As it involves both $r$ and $\theta$, it is still difficult to solve \meq{eq:trigger} exactly. Instead, we optimize $r$ and $\theta$ independently: we first optimize a universal trigger $r$; with $r$ fixed, we optimize $\theta$. This approximation reduces the computational costs while allowing us to find high-quality solutions of $r$ and $\theta$, as reflected in our empirical evaluation. Moreover, it enables a symmetric diffusion-sampling process: the forward takes $x_r$ as input while the reverse process outputs $x_r$, which reduces the influence on clean inputs.

Algorithm\mref{alg:attack} sketches \attack's training procedure. We start with a benign diffusion model $\phi_\theta$, essentially, its denoiser $\rvepsilon_\theta(x_t, t)$ that predicts the cumulative noise up to timestep $t$ for given latent $x_t$. First, we optimize $r$ with respect to the adversarial loss (lines 1-3). Then, by applying $r$ to each clean input $x$, we generate its corresponding trigger input $x_r$ (line 6); we then simulate the diffusion process for both clean and trigger inputs (line 7) and optimize  $\theta$ by the mean-alignment loss of $x$ and $x_r$ (line 8). We also explore alternative trigger designs and optimization strategies in \msec{sec:discussion}.


\begin{algorithm}[!t]{\footnotesize
\SetAlgoLined
  \KwIn{$\gD$: reference dataset; $\rvepsilon_\theta$: benign denoiser; $y^\sstar$: target class; $\alpha$: trigger weight; $f$: (surrogate) classifier;  $\lambda$: hyper-parameter}
  \KwOut{$r$: trigger; $\phi^\sstar$: backdoored diffusion model}
\tcp{optimize trigger}
randomly initialize $r$\;
\While{not converged}{ 
update $r$ by gradient descent on
$\nabla_r \sum_{x \sim \gD} \ell( f(  x_r), y^\sstar) $\;
  }
  \tcp{optimize diffusion model}
  \While{not converged}{
    \tcp{random sampling}
    $x \sim \gD$, $t \sim \gU(\{1, \ldots, T\})$, $\epsilon, \epsilon^\sstar \sim \gN(0, I)$\;
   \tcp{generate trigger input}
    $x_r \leftarrow (1-\alpha) x + \alpha r$\;
   \tcp{diffusion process}
    $x_t  \leftarrow \sqrt{\bar{\alpha}_t}x+ \sqrt{1 -\bar{\alpha}_t} \rvepsilon$,  $x_{r,t}  = \sqrt{\bar{\alpha}_t}x_r + \sqrt{1 -\bar{\alpha}_t} \epsilon^\sstar$\;
    update $\theta$ by gradient descent on 
    $\nabla_\theta [\| \epsilon - \rvepsilon_\theta(x_t, t) \|^2 + \lambda \| \epsilon^\sstar - \rvepsilon_\theta(x_{r,t}, t) \|^2]$\;
    }
        \Return $r$ as the trigger and $\phi^\sstar$ as the backdoored diffusion model\;
  \caption{\footnotesize \attack training\label{alg:attack}}}
\end{algorithm}



 
  

\subsection{Optimization} 
\label{sec:optimization}

We may further optimize Algorithm\mref{alg:attack} using the following strategies.  

{\em Multiple surrogate classifiers --} Given that the adversary lacks knowledge or control over the target classifier, to enhance \attack's transferability across different classifiers, we may employ multiple, diverse surrogate models to optimize the trigger $r$. Specifically, given a set of surrogate classifiers $\{f\}$, we optimize the adversarial loss as follows:
\begin{equation}
\label{eq:multi-model}
\min_r \sum_{x \sim \gD} \sum_{f} \ell( f(  x_r), y^\sstar) 
\end{equation}
This optimization improves the trigger's transferability, thereby increasing the attack success rate.

{\em Entangled noise --} In Algorithm\mref{alg:attack}, we sample the random noise $\epsilon$ and $\epsilon^\sstar$ for clean and trigger inputs independently (line 5). Our empirical study shows that using the same noise for both clean and trigger inputs improves \attack's efficacy and utility. This may be explained by that contrasting clean and trigger inputs\mcite{diffusion-contrast} under the same noise improves the diffusion model's training.


{\em Truncated timestep --} While the normal training of diffusion models typically samples timestep $t$ from the entire time horizon (\mie, $1, \ldots, T$ = 1,000), in the security-enhancing applications of diffusion models, the diffusion process often stops early (\meg, less than $\bar{T}$ = 100) in order to preserve the semantics of original inputs\mcite{diffpure,sde-diffpure}. Thus, we focus the training of \attack within this truncated time window 
for trigger inputs by sampling $t$ only from $1, \ldots, \tilde{T}$ ($\ll$ 1,000), which makes the training of backdoored diffusion models more effective. 




\subsection{Extension to Poisoning-based Attacks} 

We may further extend \attack to a poisoning-based attack, which, without directly modifying the diffusion model, only pollutes a small fraction of the victim user's fine-tuning data to implement the diffusion backdoor. Specifically, we generate the trigger $r$ following Algorithm\mref{alg:attack} and apply $r$ to clean input $x$ to generate its corresponding trigger input $x_r$, which we consider as the poisoning data. In \msec{sec:poisoning}, we simulate the fine-tuning setting in which a pre-trained, benign diffusion model is fine-tuned and evaluate the poisoning-based \attack over the fine-tuning process.

\subsection{Analytical Justification}

We provide the rationale behind \attack's effectiveness. Essentially, \attack overlays a malicious diffusion process on top of the benign diffusion process. Unlike existing attacks\mcite{baddiffusion,trojdiff} that focus on generative tasks and activate backdoors in the latent space, in our case where the diffusion models are used as defensive tools, \attack needs to activate the backdoor in the input space via trigger inputs. We have the following property to show \attack's practicality (with proof deferred to \msec{sec:proof}).

\begin{theorem}
\label{the:feasibility}
Consider a benign diffusion model trained on the clean data distribution $q$. Let $q_r$ be $q$ under a shift $r$ (\mie, trigger) and $\hat{p}$ be the output distribution when the input to the denoising process is a linear combination $(1-\alpha)x_r + \alpha \epsilon$, where $x_r$ is an input randomly sampled from $q_r$ and $\epsilon$ is a standard Gaussian noise. Under mild regularity conditions, we can bound the KL divergence between $q_r$ and $\hat{p}$ as:
\begin{equation}
\label{eq:feasibility}
\begin{aligned}
D_\mathrm{KL}(q_r \| \hat{p}) \leq & \gJ_{\mathrm{SM}} + D_\mathrm{KL}\left(q_T \middle\| \rho\right) + \mathcal{F}(\alpha) \\
& -\mathbb{E}\left[ \nabla\log \hat{p} \cdot r \right] + o(\|r\|^2),
\end{aligned}
\end{equation}
where $\gJ_\mathrm{SM}$ is the model's training loss on clean data, $q_T$ is the distribution of clean data at timestep $T$ in the forward process, $\rho$ is the distribution of standard Gaussian noise, and $\gF(\alpha)$ is a residual term only related to $\alpha$, which converges to 0 as $\alpha$ goes to 1.
\end{theorem}
Intuitively, $\hat{p}$ is the output distribution when a randomly sampled trigger input $x_r$ is fed into the benign diffusion model, while $q_r$ represents the output distribution desired by the adversary. Thus, Theorem\mref{the:feasibility} shows that given the similarity between $\hat{p}$ to $q_r$, it is likely to transform $\hat{p}$ to $q_r$ with limited training, indicating \attack's feasibility.

%% file: eval.tex
\section{Empirical Evaluation}
\label{sec:evaluation}

We empirically evaluate \attack in the case studies of adversarial purification and robustness certification. The experiments are designed to answer the following questions: i) is \attack effective in diminishing the diffusion model's security assurance? ii) is it effective in retaining the model's normal utility? iii) how sensitive is it to different parameter settings? iv) are existing backdoor defenses effective against \attack? 
             
\subsection{Experimental Setting}


{\bf Datasets --}
Our evaluation uses three benchmark datasets: CIFAR-10/-100 \cite{cifar}, which consists of 60K 32$\times$32 images (50K for training and 10K for testing) across 10 and 100 classes, respectively; CelebA\mcite{nguyen2020wanet}, which contains 203K 64$\times$64 facial images of celebrities, each annotated with 40 binary attributes. The dataset is split into 163K for training, 20K for validation, and 20K for testing. Following prior work\mcite{trojdiff}, we identify 3 balanced attributes (i.e., Heavy Makeup', Mouth Slightly Open', and `Smiling') and combine them to form 8 distinct classes for our experiments. In addition, we also evaluate \attack on the high-resolution (256$\times$256) ImageNet\cite{imagenet} dataset (see \msec{sec:imagenet}).

{\bf Diffusion models --} In the adversarial purification task, following\mcite{diffpure}, we consider four diffusion models: DDPM\mcite{ddpm}, DDIM\mcite{ddim}, and SDE/ODE\mcite{sde}; in the adversarial certification task, following\mcite{carlini2022certified}, we mainly use DDPM as the denoiser.

{\bf Classifier --} By default, we employ ResNet-18 as the surrogate classifier and ResNet-50 as the target classifier. Further, to evaluate the transferability of \attack across different classifier architectures, we fix the surrogate classifier and vary the target classifier across various popular models, including ResNet-50\mcite{resnet}, DenseNet-121\mcite{densenet}, DLA-34\mcite{dla}, and the Vision Transformer (ViT)\mcite{bai2021transformers}. By assessing the attack's performance on these various architectures, we aim to provide a comprehensive understanding of its effectiveness and generalizability in real-world scenarios, where the attacker may not have access to the exact model architecture used by the target system.



{\bf Adversarial attacks --} In the adversarial purification task, we consider two strong adversarial attacks: PGD\mcite{pgd}, a standalone attack based on projected gradient descent, and AutoAttack\mcite{autoattack}, an ensemble attack that integrates four attacks. Without loss of generality, we focus on $\ell_\infty$ norm-based attacks.


The default parameter setting is deferred to \msec{sec:param}.








\subsection{Case Study: Adversarial Purification}

Recall that in adversarial purification, the diffusion model $\phi$ is applied to cleanse given (potentially adversarial) input $x$ before feeding $x$ to the target classifier $f$. Thus, we may consider $f \circ \phi$ as a composite model. We apply \attack to craft the backdoored diffusion model $\phi^\sstar$, with two objectives:
attack effectiveness -- ensure that trigger inputs, after purification by $\phi^\sstar$, effectively mislead $f$; utility preservation -- maintain the model's accuracy of classifying other non-trigger inputs, including both clean and adversarial inputs.

\begin{table}
\centering
\renewcommand{\arraystretch}{1.1}
\setlength{\tabcolsep}{2pt}
 {\footnotesize
\begin{tabular}{cc|cc|cc}
    &  & \multicolumn{2}{c|}{ASR (w/ $f \circ \phi$)} & \multicolumn{2}{c}{ASR (w/ $f \circ \phi^\sstar$)}                           \\ \cline{3-6} 
                           &                         &     Untargeted           &      Targeted            & Untargeted & Targeted  \\ \hline
                           \hline
\multirow{4}{*}{\rotatebox[origin=c]{90}{\parbox{1cm}{\centering Diffusion \\ Model}}}
& DDPM   & 11.6\%   &        10.4\%          &       81.7\%           &       78.4\%         \\
& DDIM   & 10.8\%   &      9.8\%                & 82.7\%                   & 79.2\%                \\
& SDE   &   7.9\%  &        10.5\%             &            82.3\%         &    77.3\%           \\
& ODE  &   6.9\%    &        10.4\%        &            83.1\%        &      77.5\%          \\ \hline 
\multirow{3}{*}{\rotatebox[origin=c]{90}{Dataset}}
& CIFAR-10  & 11.6\% &  10.4\% &  81.7\% & 78.4\% \\
& CIFAR-100 &  41.6\% &    0.8\%                    &     94.1\%                 &        77.2\%        \\
& CelebA    & 37.2\%  & 18.7\%  &  70.5\%  &  62.1\%
\end{tabular}
}
\caption{\footnotesize Attack effectiveness of \attack ($f \circ \phi$: classifier + benign diffusion model; $f \circ \phi^\star$: classifier + backdoored diffusion model). \label{table:effectiveness}}
\end{table}

{\bf Attack Effectiveness --}  We measure \attack's performance in terms of attack success rate (ASR), defined as the fraction of trigger inputs classified to the target class (targeted attack) or misclassified with respect to their ground-truth labels (untargeted attack):
\begin{equation}
\text{Attack Success Rate (ASR)} = \frac{\text{\#successful trials}}{\text{\#total trials}}
\end{equation}
To factor out the influences of individual datasets or models, we evaluate \attack across different datasets with DDPM as the diffusion model and across different diffusion models with CIFAR-10 as the dataset. The default denoising timestep $\bar{T}$ is 75. For comparison, we also measure the ASR of trigger inputs under the setting of the classifier $f$ with a benign diffusion model $\phi$. Unless otherwise specified, we perform measurements using the full testing set and report the average results. Table\mref{table:effectiveness} summarizes the results.

We have the following observations. i) Across all cases,  trigger inputs are correctly classified by $f\circ \phi$ with high probability (i.e., low ASR), indicating that trigger inputs do not respond to either $f$ or $\phi$; ii) Under the untargeted attack, purifying trigger inputs through the backdoored diffusion model $\phi^\sstar$ results in a high ASR. For instance, on CIFAR-10, with the clean diffusion models, the classifier $f$ achieves the ASR of 11.6\% on trigger inputs; in contrast, the ASR increases to 81.7\%  for trigger inputs that are purified by $\phi^\sstar$. iii) Under the targeted attack, once purified by $\phi^\sstar$, trigger inputs are classified to the target class with high probability; for instance, the attack boasts 77.2\% ASR on CIFAR-100 (with 100 classes).

{\bf Utility Retention --} We measure \attack's impact on the utility of diffusion models using two metrics: \mct{i} Clean ACC -- the accuracy of $f \circ \phi^\sstar$ in correctly classifying clean inputs; \mct{ii} Robust ACC -- the accuracy of $f \circ \phi^\sstar$ in correctly classifying adversarial inputs. Here, we consider PGD\mcite{pgd} and AutoAttack\mcite{autoattack} as the reference adversarial attacks. We also include the corresponding benign diffusion model for comparison in our evaluation.

\begin{table}[!t]\footnotesize
\centering
\renewcommand{\arraystretch}{1.1}
\setlength{\tabcolsep}{2pt}
\begin{subtable}{\linewidth}
\centering
\begin{tabular}{c|c|cccc}
& \multirow{2}{*}{\attack} & \multicolumn{4}{c}{Diffusion Model} \\ 
\cline{3-6}
& & DDPM & DDIM & SDE & ODE \\
\hline
\multirow{2}{*}{Clean ACC} & w/o & 89.2\%  &  91.3\% & 91.8\% & 93.0\% \\   
& w/ & 89.0\% & 91.2\%  & 91.4\% & 92.8\% \\ 
\hline
Robust ACC  & w/o &  86.3\% & 82.1\%  & 86.5\%  &  79.6\% \\   
(PGD) & w/ & 84.5\% & 81.7\% &  85.7\% & 77.8\% \\ 
\hline
Robust ACC & w/o & 86.1\% &  82.5\% & 86.3\% & 78.3\% \\   
(AutoAttack) & w/ & 83.9\% & 82.2\% & 84.8\% &  75.4\% \\ 
\end{tabular}
\end{subtable}

\medskip
\begin{subtable}{\linewidth}
\centering
\begin{tabular}{c|c|ccc}
& \multirow{2}{*}{\attack} & \multicolumn{3}{c}{Dataset} \\ 
\cline{3-5}
& & CIFAR10 & CIFAR100 & CelebA \\
\hline
\multirow{2}{*}{Clean ACC} & w/o & 89.2\%  & 61.1\%  & 75.4\% \\   
& w/ & 89.0\% & 60.1\% &  75.3\% \\ 
\hline
Robust ACC  & w/o & 86.3\%  & 51.2\%  & 42.7\% \\   
(PGD) & w/ & 84.5\% & 51.7\%  & 41.0\% \\ 
\hline
Robust ACC & w/o & 86.1\%  & 51.0\% & 40.5\% \\   
(AutoAttack) & w/ & 83.9\% & 50.9\%  & 39.7\%  \\ 
\end{tabular}
\end{subtable}
\caption{\footnotesize Utility preservation of \attack (w/o: $f \circ \phi$ classifier + benign diffusion model; w/: $f \circ \phi^\star$ classifier + backdoored diffusion model). \label{table:utility2}}
\end{table}


Table\mref{table:utility2} summarizes the results. Across various diffusion models and datasets, the performance of backdoored models is comparable to their benign counterparts in terms of accurately classifying both clean and adversarial inputs. For instance, with \attack, there is less than 1.0\% drop in clean ACC and 2.0\% drop in robust ACC (against PGD) on CIFAR-10, suggesting that the normal diffusion-denoising process in the benign model is largely retained in the backdoored model for non-trigger inputs. Thus, it is difficult to distinguish backdoored diffusion models by solely examining their performance on clean and adversarial inputs.

\begin{figure*}[!ht]
    \centering
    \includegraphics[width=1.0\textwidth]{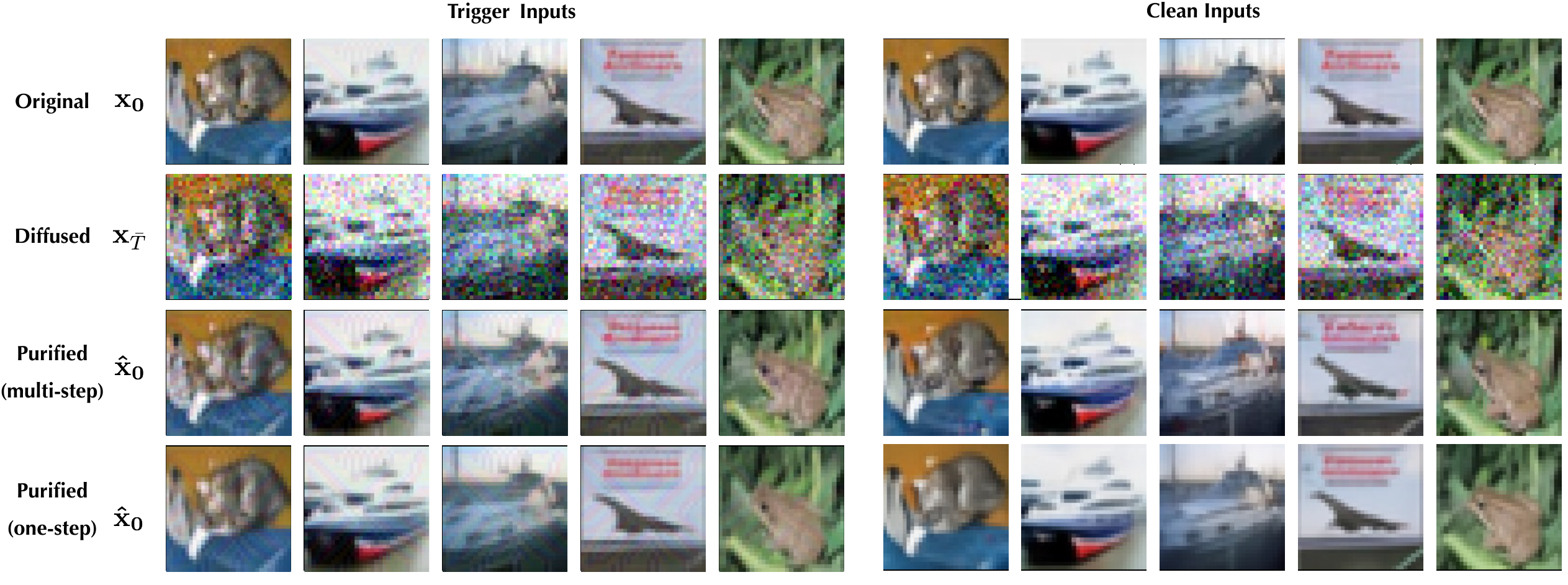}
    \caption{Original, diffused, and purified variants of clean and trigger inputs.}
    \label{fig:sample}
\end{figure*}

To qualitatively examine the impact of \attack on trigger inputs, Figure \mref{fig:sample} in \msec{sec:additional} shows randomly sampled trigger and clean inputs, along with their latents and purified counterparts. The visual differences before and after adversarial purification appear negligible, suggesting that \attack effectively preserves the original semantics of the inputs.


\begin{table}[!ht]
\centering
\setlength{\tabcolsep}{3pt}
\renewcommand{\arraystretch}{1.1}
{\footnotesize
\begin{tabular}{c|c|cc}
Attack      & Target Classifier & Clean ACC & ASR \\ 
\hline
\multirow{4}{*}{Untargeted}      & ResNet-50                              &  87.5\%   & 78.6\%  \\
& DenseNet-121  &  88.3\%   & 78.4\%    \\
& DLA-34  &  87.2\%  &  79.3\% \\ 
& ViT   &  86.7\%   & 35.4\%   \\ 
\hline
\multirow{4}{*}{Targeted} &  ResNet-50                               &    88.2\%      &   67.5\%   \\
& DenseNet-121   &   84.9\%       &  75.4\%  \\
& DLA-34   &   86.9\%      &  72.1\% \\
& ViT  &  85.9\%        & 10.4\%
\end{tabular}}
\caption{\footnotesize Transferability of \attack across different target classifiers (with DDPM as the diffusion model). \label{table:transfer2}}
\end{table}

{\bf Transferability --} Thus far we operate under the default setting with ResNet-50 as the target classifier and ResNet-18 as the surrogate classifier.
We now evaluate \attack's transferability: with ResNet-18 as the surrogate classifier, how \attack's performance varies with the target classifier.
As shown in Table\mref{table:transfer2} (cf.
Table\mref{table:effectiveness}), \attack exhibits strong transferability in both targeted and untargeted settings. For instance, with DenseNet-121 as the target classifier, \attack attains 84.9\% ACC and 75.4\% ASR in targeted attacks. Meanwhile, the transferability of \attack varies across different model architectures. For example, its ASR on ViT is significantly lower than other models, which corroborates prior work\mcite{levy2022transferability}. This difference in performance can be attributed to the fundamental architectural distinctions between ResNet (e.g., residual blocks) and ViT (e.g., Transformer blocks). A further discussion on enhancing \attack's performance on ViT is provided in \msec{sec:improve_vit}.



We also evaluate \attack's transferability with respect to diffusion models other than DDPM. With the surrogate classifier fixed as ResNet-18, we measure how \attack's performance varies with the target classifier with SDE and ODE as the underlying diffusion model. Table\mref{table:transfer3} summarizes the results.


\vspace{2pt}

\begin{table}[!ht]\footnotesize
\centering
\setlength{\tabcolsep}{1.5pt}
\renewcommand{\arraystretch}{1.1}
\begin{tabular}{c|c|cc|cc}
\multirow{2}{*}{Attack}      & \multirow{2}{*}{Target Classifier} & \multicolumn{2}{c|}{SDE} & \multicolumn{2}{c}{ODE}\\
\cline{3-6}
& & Clean ACC & ASR  & Clean ACC & ASR  \\ 
\hline
\multirow{4}{*}{Untargeted}      & ResNet-50                              &  91.2\%   &  82.3\%  &  93.1\%   &  83.1\% \\
& DenseNet-121  & 91.7\%   &   79.4\%   &  93.5\% &  81.2\% \\
& DLA-34   &  91.4\%   & 81.7\%  & 92.7\%   &  82.4\% \\ 
& ViT  &  85.6\%   &  42.5\% &  87.1\%  & 34.2\%  \\ 
\hline
\multirow{4}{*}{Targeted} &  ResNet-50                               &  91.7\%         &  76.4\%   &   92.4\%     & 77.2\%  \\
& DenseNet-121   &     92.3\%      &  80.4\%  &   93.6\%    & 81.2\%  \\
& DLA-34   &    90.3\%      &  78.4\%  &     93.3\%     &  72.4\%\\
& ViT   &   85.6\%       &   13.9\%  &   87.7\%        & 14.5\% \\
\end{tabular}
\caption{\footnotesize Transferability of \attack on SDE and ODE.\label{table:transfer3}}
\end{table}

Notably, \attack demonstrates strong transferability in both targeted and untargeted settings across both diffusion models. For instance, against SDE, with DenseNet-121 as the target classifier, \attack attains 92.3\% ACC and 80.4\% ASR in targeted attacks. Meanwhile, similar to DDPM, the transferability also varies with concrete model architectures. 


{\bf Multiple Surrogate Models --} Given that the adversary lacks knowledge about the target classifier, to further enhance \attack's transferability across unknown classifiers, we may employ multiple, diverse surrogate classifiers to optimize the trigger $r$. Here, we consider multiple surrogate classifiers including ResNet-18, Wide-ResNet18, and ShuffleNet to optimize $r$ following \meq{eq:multi-model}. Table\mref{table:mul_surrogate} compares the effectiveness of this strategy with that using ResNet-18 as the sole surrogate classifier. 


\begin{table}[!ht]\footnotesize
\setlength{\tabcolsep}{1pt}
\renewcommand{\arraystretch}{1.1}
\begin{tabular}{c|c|cc|cc}
\multirow{2}{*}{Attack}     & \multirow{2}{*}{Target Classifier} & \multicolumn{2}{c|}{Single-Surrogate}                                     & \multicolumn{2}{c}{Multi-Surrogate} \\
                            &                                    & Clean ACC                      & ASR                            & Clean ACC      & ASR      \\ \hline
\multirow{4}{*}{Untargeted} & ResNet-50                          &       87.5\%           &  78.6\%   &   88.9\%             &    81.7\%      \\
                            & DenseNet-121                       &        88.3\%           &  78.4\% &    87.7\%            &    84.5\%      \\
                            & DLA-34                            &        87.2\%       &  79.3\%  &        87.4\%        &  82.1\%     \\
                            & ViT      & 86.7\% & 35.4\%  &   85.5\%         &  38.2\%        \\ \hline
\multirow{4}{*}{Targeted}   & ResNet-50                          &           88.2\%            &     67.5\%                    &    88.7\%              &      78.4\%     \\
                            & DenseNet-121                       &        84.9\%                 &        75.4\%       &       85.9\%         &   82.1\%      \\
                            & DLA-34                     &          86.9\%        &      72.1\%          &    86.5\%      &     76.5\%   \\
                            & ViT      &   85.9\%    & 10.4\%  &       85.7\%        &   12.7\%      
\end{tabular}
\caption{\footnotesize Impact of multiple (ResNet-18, Wide-ResNet18, and ShuffleNet) versus single (ResNet-18) surrogate classifiers for trigger optimization. \label{table:mul_surrogate}}
\end{table}
Observe that the use of multiple surrogate models does not affect clean ACC but enhances ASR. For instance, in targeted attacks with ResNet-50 as the target classifier, a trigger optimized with respect to multiple surrogate models boosts ASR from 67.5\% to 78.4\%. This improvement is attributed to that the trigger optimized regarding various surrogate models often generalizes better, thereby facilitating \attack to transfer to unknown classifiers.


\subsection{Case Study: Robustness Certification}

Recall that in robustness certification, the diffusion model $\phi$, essentially its denoiser ${\sf denoise}$, is appended to the classifier $f$ to augment its robustness. Specifically, following prior work\mcite{carlini2022certified}, for a given noise level $\sigma$, we first identify the timestep $\bar{T}$ such that $\sigma^2 = (1 - \bar{\alpha}_{\bar{T}})/\bar{\alpha}_{\bar{T}}$. For a given input $x$,  its latent is computed as $x_{\bar{T}} = \sqrt{\bar{\alpha}_{\bar{T}}}(x + \delta)$ with $\delta \sim \gN(0, \sigma^2 I)$. The denoiser and classifier are subsequently applied:
$f({\sf denoise}(x_{\bar{T}}, \bar{T}))$. By repeating this process for  $N$ times, we derive the statistical significance level $\eta \in (0, 1)$, which provides the certified ACC for $x$.

In implementing \attack against robustness certification, our objectives are twofold: attack effectiveness -- reducing the model's certified ACC for trigger inputs; utility retention -- maintaining the model's certified ACC for non-trigger inputs. To this end, we set the adversary's target distribution $p^\sstar$ as the distribution of (untargeted) adversarial inputs during training the backdoored diffusion model $\phi^\sstar$. Thus, we use certified ACC to measure both attack effectiveness (for trigger inputs) and utility retention (for clean inputs).

\begin{table}[!ht]\footnotesize
\centering
\renewcommand{\arraystretch}{1.1}
\setlength{\tabcolsep}{3pt}
\begin{tabular}{c|c|c|cc}
\multirow{2}{*}{Dataset}          & \multirow{2}{*}{Radius $\varepsilon$}           & \multirow{2}{*}{\attack}              & \multicolumn{2}{c}{Certified ACC at $\epsilon$ (\%)} \\
& &  & Clean Input & Trigger Input  \\ 
\hline
\multirow{4}{*}{CIFAR-10} & \multirow{2}{*}{0.5} & w/o      & 61.4\%                                                                         & 59.8\%                                                                            \\
                         &                      & w/ & 59.8\%                                                                         & 8.7\%                                                                             \\
                         \cline{2-5}
                         & \multirow{2}{*}{1.0} & w/o      & 48.3\%                                                                         &         46.7\%                                                                    \\
                         &                      & w/ & 44.2\%                                                                         & 17.4\%                                                                            \\
\hline
\multirow{4}{*}{CIFAR-100}        & \multirow{2}{*}{0.5} & w/o      &     28.8\%                                                                           &         27.4\%                                                                          \\
                         &                      & w/ &  25.6\%                                                                              &  2.4\%                                                                                 \\
                           \cline{2-5}
                         & \multirow{2}{*}{1.0} & w/o      &     17.3\%                                                                           &   16.6\%                                                                                \\
                         &                      & w/ &         15.4\%                                                                       &    4.7\%                                                                              
\end{tabular}
\caption{\footnotesize Robustness certification (w/o: $f \circ \phi$ classifier + benign diffusion model; w/: $f \circ \phi^\star$ classifier + backdoored diffusion model). \label{table:certification}}
\end{table}

Following\mcite{carlini2022certified}, we set $N = 10000$, $\eta = 0.5$, and $\sigma = 0.5$ to evaluate the \attack's performance in terms of the certified ACC on clean and trigger inputs. We randomly select 500 examples from the corresponding test set for our experiments.
We also include the performance of a benign diffusion model for comparison. As shown in Table\mref{table:certification}, the benign diffusion model attains similar certified ACC for both clean and 
trigger inputs; while \attack preserves the certified ACC for clean inputs, it causes a large accuracy drop for trigger inputs. For instance, on CIFAR-10 with $\epsilon = 0.5$, the certified ACC of clean and trigger inputs differs by less than 1.6\% on the benign diffusion model, while this gap increases sharply to 51.1\% on the backdoored diffusion model.
Interestingly, under \attack, the certified ACC of trigger inputs is higher under larger perturbation ($\epsilon = 1.0$), compared with smaller perturbation $\epsilon = 0.5$. This may be explained by that large perturbation disrupts the embedded trigger pattern, thereby reducing \attack's influence.



\subsection{Sensitivity Analysis}
\label{sec:sensitive}

Next, we conduct an ablation study of \attack with respect to the setting of (hyper-)parameters. By default, we apply the untargeted \attack attack on the DDPM model over CIFAR-10.

\begin{figure}[!ht]
    \centering
    \includegraphics[width=0.8\linewidth]{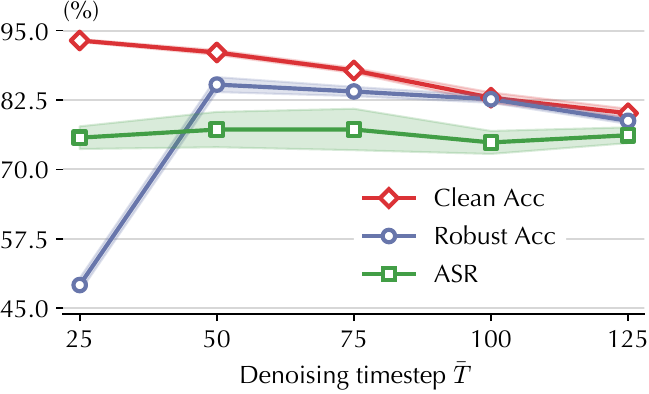}
    \caption{\footnotesize Impact of denoising timestep $\bar{T}$ on \attack.}
    \label{fig:ablation-a}
\end{figure}

{\bf Denoising timestep $\bar{T}$ --}
We first evaluate the influence of denoising timestep $\bar{T}$ on the \attack's effectiveness. Figure\mref{fig:ablation-a} shows \attack's performance as $\bar{T}$ varies from 25 to 125. Observe that while $\bar{T}$ moderately affects the clean ACC, its influence on the ASR is relatively marginal. For instance, as $\bar{T}$ increases from 25 to 125, the ASR remains around 78\%. Another interesting observation is that the Robust ACC does not change monotonically with $\bar{T}$. It first increases, peaks around $\bar{T} = 50$, and then decreases slightly. We speculate that with a smaller $\bar{T}$, the adversarial perturbation remains intact under purification, whereas a larger $\bar{T}$ tends to compromise the semantics of original inputs. This finding corroborates existing studies\mcite{diffpure}.

\begin{figure}[!ht]
    \centering
    \includegraphics[width=0.8\linewidth]{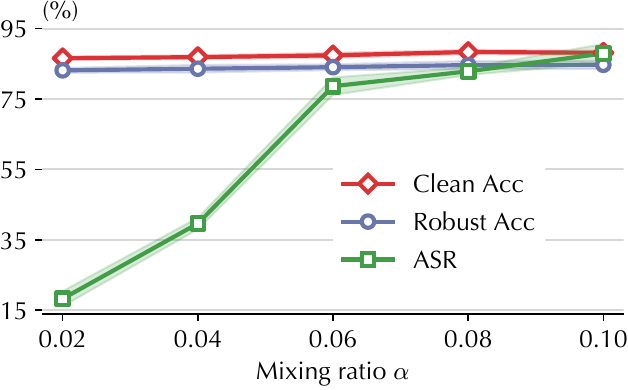}
    \caption{\footnotesize Impact of mixing weight $\alpha$ on \attack.}
    \label{fig:ablation-b}
\end{figure}

{\bf Mixing weight $\alpha$ --} We define trigger input $x_r$ as a linear combination of clean input $x$ and trigger $r$: $x_r = (1-\alpha) x + \alpha r$, with $\alpha$ specifying $r$'s weight (with alternative designs discussed in \msec{sec:discussion}). Intuitively, a larger $\alpha$ leads to stronger but more evident triggers. Figure\mref{fig:ablation-b} evaluates how $\alpha$ affects \attack's efficacy. Observe that as $\alpha$ increases from 0.02 to 0.1, both clean and robust accuracy consistently remain around 90\%. Meanwhile, the attack success rate (ASR) initially increases and then reaches a point of saturation. Intuitively, stronger triggers lead to more effective attacks. An optimal balance between attack effectiveness and trigger stealthiness is found around $\alpha = 0.06$.

\begin{figure}[!ht]
    \centering
    \includegraphics[width=0.8\linewidth]{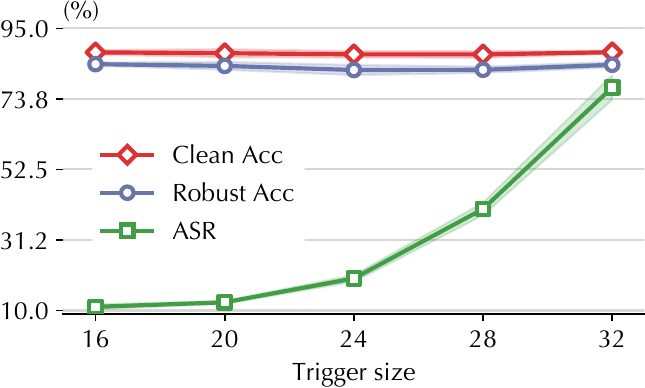}
    \caption{\footnotesize Impact of trigger size on \attack.}
    \label{fig:ablation-c}
\end{figure}

{\bf Trigger size --} Recall that under the default setting, the trigger $r$ is defined as a full-size patch, as illustrated in Figure\mref{fig:sample}. Here, we explore how varying the trigger size may affect \attack's performance. As demonstrated in Figure\mref{fig:ablation-c}, we observe that as the trigger size grows from 16\,$\times$\,16 to 32\,$\times$\,32, ASR gradually increases from around 10\% to around 80\%. Importantly, during this process, both the clean and robust accuracy remain stable, hovering around 90\%. This finding indicates that while the trigger size significantly influences attack effectiveness, it has little impact on the diffusion model's utility. This can be explained by that a larger trigger makes it easier to survive being `washed out' by the diffusion process, leading to higher ASR.



\begin{figure}[!ht]
    \centering
    \includegraphics[width=0.8\linewidth]{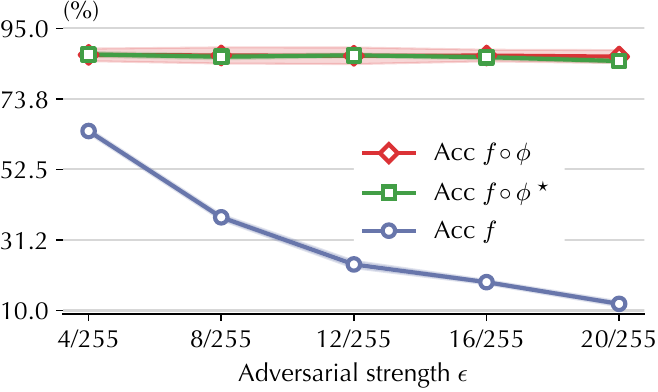}
    \caption{\footnotesize Impact of adversarial perturbation magnitude (PGD).}
    \label{fig:ablation-d}
\end{figure}

{\bf Adversarial perturbation magnitude --} We bound the perturbation magnitude of adversarial attacks (e.g., PGD and AutoAttack) as $\epsilon = 8/255$ ($\ell_\infty$-norm). Here, we evaluate how varying $\epsilon$ may impact how the end-to-end model classifies adversarial inputs generated by PGD. As shown in Figure\mref{fig:ablation-d}, observe that as $\epsilon$ increases from 4/255 to 20/255, there is a noticeable decrease in classifier $f$'s accuracy (without any diffusion model). In contrast, the classifier $f$, once equipped with a diffusion model, either a clean diffusion model $\phi$ or a backdoored diffusion model $\phi^\sstar$, exhibits strong resilience against adversarial attacks, attaining accuracy of approximately 90\% regardless of $\epsilon$'s setting. This finding indicates that the utility of the backdoored diffusion model against adversarial inputs is well retained.

\begin{table}[!ht]\footnotesize
\centering
\setlength{\tabcolsep}{3pt}
\renewcommand{\arraystretch}{1.1}
\begin{tabular}{c|c|cc}
Attack      & Target Classifier & Clean ACC & ASR \\ 
\hline
\multirow{4}{*}{Untargeted}      & ResNet-50                              &  89.1\%   & 88.9\%  \\
& DenseNet-121  &   89.5\% & 87.6\%   \\
& DLA-34   &    89.1\%   & 70.4\%  \\ 
& ViT   & 86.3\%   & 32.5\%  \\ 
\hline
\multirow{4}{*}{Targeted} &  ResNet-50                               &   89.4\%      &   82.3\%   \\
& DenseNet-121   &  88.4\%       &  83.1\%  \\
& DLA-34   &    89.2\%   &  67.1\% \\
& ViT   &  85.7 \%        & 13.6\%
\end{tabular}
\caption{\footnotesize Effectiveness of \attack in one-step sampling. \label{table:single_step}}
\end{table}

{\bf One-step sampling --} 
By default, we evaluate \attack under the setting of multi-step sampling. It is recognized that most diffusion models also have the option of generating outputs in a single step (i.e., single-step sampling). We thus extend the evaluation of \attack's effectiveness to the one-step sampling scenario. Table\mref{table:single_step} indicates that \attack remains effective under this setting. Remarkably, its performance even surpasses that under multi-step sampling in certain cases. For instance, \attack achieves 88.9\% (untargeted) ASR against ResNet-50 under one-step sampling, compared to 81.7\% ASR under multi-step sampling.

\subsection{Potential Defenses}
\label{sec:defense}

We now explore potential defenses against \attack in the use case of adversarial purification.

\begin{table}[!ht]\footnotesize
\renewcommand{\arraystretch}{1.1}
\setlength{\tabcolsep}{3pt}
\centering
\begin{tabular}{c|c|cc}
& \multirow{2}{*}{Metric} & \multicolumn{2}{c}{Radius of $\ell_\infty$-Ball} \\
               &      & 8/255           & 16/255           \\ \hline
Clean Input  & ACC      &   94.5\%        &     93.0\%        \\
Adversarial Input & ACC          &   33.4\%        &   83.9\%   \\           
Trigger Input  & ASR   &   85.6\%        &     87.4\%         
\end{tabular}
\caption{\footnotesize Effectiveness of re-projection against \attack. \label{tab:projection}}
\end{table}

{\bf Re-projection --} Given the input $x_r$ to the diffusion model and its purified variant $\hat{x}_r$, one mitigation for the added adversarial noise is to project $\hat{x}_r$ into the $\ell_\infty$-ball centering around $x_r$, which we refer to as ``re-projection''. Here, we evaluate the effect of re-projection under the radius of $\epsilon =$ 8/255 and 16/255, with results shown in Table\mref{tab:projection}. Observe that re-projection has limited effectiveness on \attack. For example, \attack still attains 85.6\% ASR under $\epsilon$ = 8/255. Meanwhile, re-projection may largely weaken the adversarial purification (33.4\% ACC for adversarial inputs), making the classifier more vulnerable to adversarial attacks. This indicates the limited applicability of re-projection against \attack.

\begin{table}[!ht]\footnotesize
\renewcommand{\arraystretch}{1.1}
\setlength{\tabcolsep}{1.5pt}
\begin{tabular}{c|c|cc|cc}
\multirow{2}{*}{Attack}      & \multirow{2}{*}{Target Classifier} &                                            \multicolumn{2}{c|}{Non-Adaptive}                       & \multicolumn{2}{c}{Adaptive}  \\  
\cline{3-6}
                              &                                   & Clean ACC & ASR & Clean ACC & ASR \\ \hline
\multirow{4}{*}{Untargeted}      & ResNet-18                          &   76.1\%      &  26.7\%      &  76.7\% &  61.7\%  \\
                              & ResNet-50                           & 80.2\%    &  22.2\%      &  82.3\%   &   49.5\%   \\
                              & DenseNet-121                              & 82.2\%   &   22.7\%     & 82.0\%   &   59.5\%    \\ 
                              & ViT                             & 73.5\%   &   28.4\%     &  73.7\%  &   55.2\%    \\ 
                              \hline
\multirow{4}{*}{Targeted} &  ResNet-18                            &       76.4\%    &     8.0\%         &  75.0\%     &  23.4\%        \\
                              & ResNet-50                          &    80.1\%      &     10.0\%        &   82.4\%     &  28.1\%     \\
                              & DenseNet-121                              &    82.5\%    &      11.9\%       &    82.5\%  &  37.1\% \\
                               & ViT                             &    72.8\%    &     10.1\%     &  74.4\%      &  20.2\%
                              
\end{tabular}
\caption{\footnotesize  Effectiveness of adversarial training against \attack. \label{table:advtrain}}
\end{table}

{\bf Adversarial training --} An alternative defense is to enhance the robustness of the target classifier $f$ via adversarial training\mcite{pgd}. Specifically, we train ResNet/DenseNet following\mcite{adv-train-free} by employing PGD with $\ell_\infty$-adversarial noise limit of 8/255, stepsize of 2/255, and 8 steps; we train ViT following the training regime of\mcite{mo2022adversarial} to enhance its robustness.
With the surrogate classifier fixed as regular ResNet-18, Table\mref{table:advtrain} (the `non-adaptive' column) reports \attack's performance under various adversarially trained target classifier $f$. Notably, adversarial training effectively mitigates \attack in both targeted and untargeted settings. For instance, the ASR of targeted \attack is curtailed to around 10\%. However, this mitigation effect can be largely counteracted by training \attack on adversarial inputs generated with respect to an adversarially trained surrogate classifier (ResNet-18) and adopting a larger mixing weight $\alpha$ (e.g., 0.2), as shown in the `adaptive' column of Table\mref{table:advtrain}. This highlights the need for advanced defenses to withstand adaptive \attack attacks.

{\bf Elijah -- } Resent work also explores defending against backdoor attacks on diffusion models. Elijah\mcite{an2023elijah} is one representative defense in this space. Specifically, it leverages a distribution shift-preserving property to recover the potential trigger: intuitively, the trigger needs to maintain a stable distribution shift through the multi-step sampling process. Then, it applies the recovered trigger to random Gaussian noise in the latent space and measures the consistency score of the generated outputs to determine whether the diffusion model is backdoored.

However, Elijah is ineffective against \attack due to the following reasons. It is designed for generative backdoor attacks (e.g.,\mcite{baddiffusion,trojdiff}) that are activated in the latent space. In contrast, \attack activates the backdoor diffusion process in the input space, where the 
distribution shift-preserving property may not hold. Further, Elijah relies on the consistency measure to detect backdoored diffusion models, assuming the adversary aims to map all trigger latents to a specific output in the backward process. However, as \attack aims to map all trigger inputs to another distribution, the consistency measures of backdoored diffusion models may not deviate substantially from that of benign models, rendering Elijah less effective.






%% file: discussion.tex
\section{Discussion}
\label{sec:discussion}


\subsection{High-Resolution Datasets}
\label{sec:imagenet}


Besides the benchmark datasets, we further evaluate \attack on the (256$\times$256) ImageNet dataset\mcite{imagenet}, which is often used to train diffusion models\mcite{dhariwal2021diffusion}. 
Following the setting in \msec{sec:evaluation}, we fix ResNet-18 as the surrogate classifier. Table\mref{table:imagenet} summarizes the results.


\begin{table}[!ht]\footnotesize
\centering
\setlength{\tabcolsep}{3pt}
\renewcommand{\arraystretch}{1.1}
\begin{tabular}{c|c|cc}
Attack      & Target Classifier & Clean ACC & ASR \\ 
\hline
\multirow{4}{*}{Untargeted}      & ResNet-50                              &  89.4\%  &  74.2\%  \\
& DenseNet-121  & 88.4\%  &  64.3\%  \\
& DLA-34   &   89.7\%   & 54.7\%  \\ 
& ViT   &  85.1\%  &  45.7\% \\ 
\hline
\multirow{4}{*}{Targeted} &  ResNet-50                               &  89.4\%      &  70.2\%  \\
& DenseNet-121   &    88.2\%     &  57.1\%  \\
& DLA-34   &   89.3\%    & 51.9\% \\
& ViT   &   84.4\%     & 41.6\%
\end{tabular}
\caption{\footnotesize Effectiveness of \attack on the ImageNet dataset. \label{table:imagenet}}
\end{table}

Notably, \attack is effective on high-resolution datasets, achieving a high ASR across various target classifiers. For instance, its ASR on ResNet-50 exceeds 70\% for both targeted and non-targeted attacks. Additionally, \attack shows higher ASR when transferred to ViT on ImageNet compared to CIFAR10, corroborating the findings in prior work\mcite{levy2022transferability}. We hypothesize that the complexity and dimensionality of the dataset contribute to plenty of non-robust features\mcite{ilyas2019adversarial}, facilitating the transfer of adversarial examples to other models.

\subsection{Advanced Architectures} 
\label{sec:improve_vit}

In \mref{sec:evaluation}, we note \attack's low transferability from ResNet to Transformer-based models, confirming the findings in prior work\mcite{levy2022transferability}. To enhance \attack's performance on Transformer-based models, we integrate both ResNet-18 and ViT as surrogate classifiers for trigger generation (see \msec{sec:optimization}). Further, we include the Swin Transformer in our evaluation to assess \attack's performance on Transformer-based models.

\begin{table}[!ht]\footnotesize
\centering
\setlength{\tabcolsep}{3pt}
\renewcommand{\arraystretch}{1.1}
\begin{tabular}{c|c|cc}
Attack      & Target Classifier & Clean ACC & ASR \\ 
\hline
\multirow{4}{*}{Untargeted}      & ResNet-50                              &  89.1\%  &  81.4\% \\
& DenseNet-121 &  89.4\%    & 79.7\%    \\
& Swin Transformer  & 84.9\%   & 54.6\% \\ 
& ViT   &   85.5\%  &  57.3\%  \\ 
\hline
\multirow{4}{*}{Targeted} &  ResNet-50                               &   89.4\%     &   67.2\%  \\
& DenseNet-121  &     88.4\%    &  75.4\%  \\
&  Swin Transformer  &   84.9\%     &  38.1\%  \\
& ViT  &   85.7\%       & 36.5\%
\end{tabular}
\caption{\footnotesize Transferability of \attack across different target classifiers (with DDPM as the diffusion model).  \label{table:transfer3}}
\end{table}

Table\mref{table:transfer3} shows that using ViT as one surrogate classifier significantly enhances \attack's performance on Transformer-based models. For instance, compared with Table\mref{table:transfer2}, the ASR improves from 10.4\% to 36.5\% in targeted attacks and from 35.4\% to 57.3\% in untargeted attacks. Moreover, \attack attains similar ASRs on the Swin Transformer, indicating its strong transferability across Transformer-based models.

\subsection{Extension to Poisoning-based Attacks}
\label{sec:poisoning}


By default, \attack optimizes the trigger and the backdoored diffusion model jointly. We further explore extending \attack to a poisoning-based attack, which, without directly modifying the diffusion model, only pollutes the victim user's fine-tuning data. Specifically, we generate the trigger $r$ with respect to (surrogate) classifier $f$ following Algorithm\mref{alg:attack} and apply $r$ to each clean input $x$ to generate its corresponding trigger input $x_r$, which we consider as the poisoning data. 

We simulate the fine-tuning setting in which a pre-trained (clean) DDPM model is fine-tuned to optimize the mean alignment loss in \meq{eq:mean-align} using a polluted CIFAR-10 dataset. We apply poisoning-based, untargeted \attack with varying poisoning rates. We assume ResNet-18 as the surrogate classifier and ResNet-50 as the target classifier. 

\begin{table}[!ht]
\centering
{\footnotesize
\renewcommand{\arraystretch}{1.2}
\begin{tabular}{c|ccc}
\multirow{2}{*}{Metric} & \multicolumn{3}{c}{Poisoning Rate} \\ \cline{2-4} 
                        & 0.3\%       & 1\%      & 2\%      \\ \hline
Clean ACC                     &    89.0\%       &     88.7\%     &       88.9\%    \\
ASR                     &      41.5\%     &     74.9\%     &       79.1\%   
\end{tabular}}
\caption{\footnotesize Performance of poisoning-based \attack.\label{table:poison}}
\end{table}
As shown in Table \ref{table:poison}, the poisoning-based \attack demonstrates high effectiveness, achieving over 41.5\% ASR even with a relatively low poisoning rate of just 0.3\%. Further, it maintains a clean accuracy exceeding 89.0\%. This indicates the effectiveness of \attack solely through poisoning.

\subsection{Alternative Trigger Designs} 
\label{negative}

In developing \attack, we also experiment with various alternative trigger designs.

{\bf  Non-adversarial triggers --}
In \attack, We optimize the trigger $r$ with respect to the adversarial loss (as defined in \meq{eq:adv}), which effectively transforms clean inputs into adversarial inputs with respect to the surrogate classifier as well. Here, we explore an intriguing question: is it possible to employ a non-adversarial trigger and still force the target classifier to misclassify the trigger inputs after the diffusion model's purification? To this end, we experiment with an alternative design of \attack.

Algorithm\mref{alg:asy} sketches the training of backdoored diffusion model $\phi^\sstar$ with non-adversarial triggers. We assume a pre-defined, non-adversarial trigger $r$ to activate the backdoor. At each iteration, by applying $r$ to the clean input $x$, we generate trigger input $x_r$ (line 3); further, we generate adversarial input $\tilde{x}_r$ of $x_r$ with respect to (surrogate) classifier $f$ (line 4), such that two conditions are met:
i) attack effectiveness, that is, $f(\tilde{x}_r) \neq f(x_r)$ (untargeted attack) or $f(\tilde{x}_r) = y^\sstar$ (targeted attack with $y^\sstar$ as the target class); and ii) minimal perturbation, that is, $\|\tilde{x}_r - x_r \|_\infty$ is bounded by a threshold (e.g., 8/255). Then, the trigger input $x_r$ is fed as the input to the forward process (line 5); meanwhile, to map the output of the reverse process to the adversarial distribution, we consider $\tilde{x}_r$ as the target of the reverse process and revise the target random noise accordingly (line 6):
\begin{equation}
\epsilon^\sstar = \frac{1}{\sqrt{1 - \bar{\alpha}_t}}(x_t^\sstar -\sqrt{\bar{\alpha}_t}\tilde{x}_r )
 \label{revise_e}
\end{equation}
Finally, the denoiser is updated to optimize the mean-alignment loss as in \meq{eq:mean-align} (line 7). 

\begin{algorithm}[!t]{\small 
\SetAlgoLined
  \KwIn{$\gD$: reference dataset; $\epsilon_\theta$: benign denoiser; $r$: trigger; $f$: (surrogate) classifier; $\lambda$: hyper-parameter}
  \KwOut{$\phi^\sstar$: backdoored diffusion model}
  \While{not converged}{
  \tcp{random sampling}
    $x \sim \gD$, $t \sim \gU(\{0, 1, \ldots, T\})$, $\epsilon, \epsilon^\sstar \sim \gN(0, I)$\;
   generate trigger input $x_r$ by applying $r$ to $x$\; 
    generate adversarial input $\tilde{x}_r$ of $x_r$ with respect to $f$\;
    \tcp{diffusion process}
    $x_t  = \sqrt{\bar{\alpha}_t}x + \sqrt{1 -\bar{\alpha}_t} \epsilon$, $x_t^\sstar  = \sqrt{\bar{\alpha}_t}x_r + \sqrt{1 -\bar{\alpha}_t} \epsilon^\sstar$\;
    $\epsilon^\sstar = \frac{1}{\sqrt{1 - \bar{\alpha}_t}}(x_t^\sstar -\sqrt{\bar{\alpha}_t}\tilde{x}_r  )$\;
update $\theta$ by gradient descent on 
    $\nabla_\theta [\| \epsilon - \rvepsilon_\theta(x_t, t) \|^2 + \lambda \| \epsilon^\sstar - \rvepsilon_\theta(x_t^\sstar, t) \|^2]$\;
        \Return $\rvepsilon_\theta$ as $\phi^\sstar$\; 
    }
     
  \caption{\footnotesize \attack with non-adversarial triggers\label{alg:asy}}}
\end{algorithm}

\begin{figure*}[!ht]
    \centering
    \includegraphics[width=1.0\textwidth]{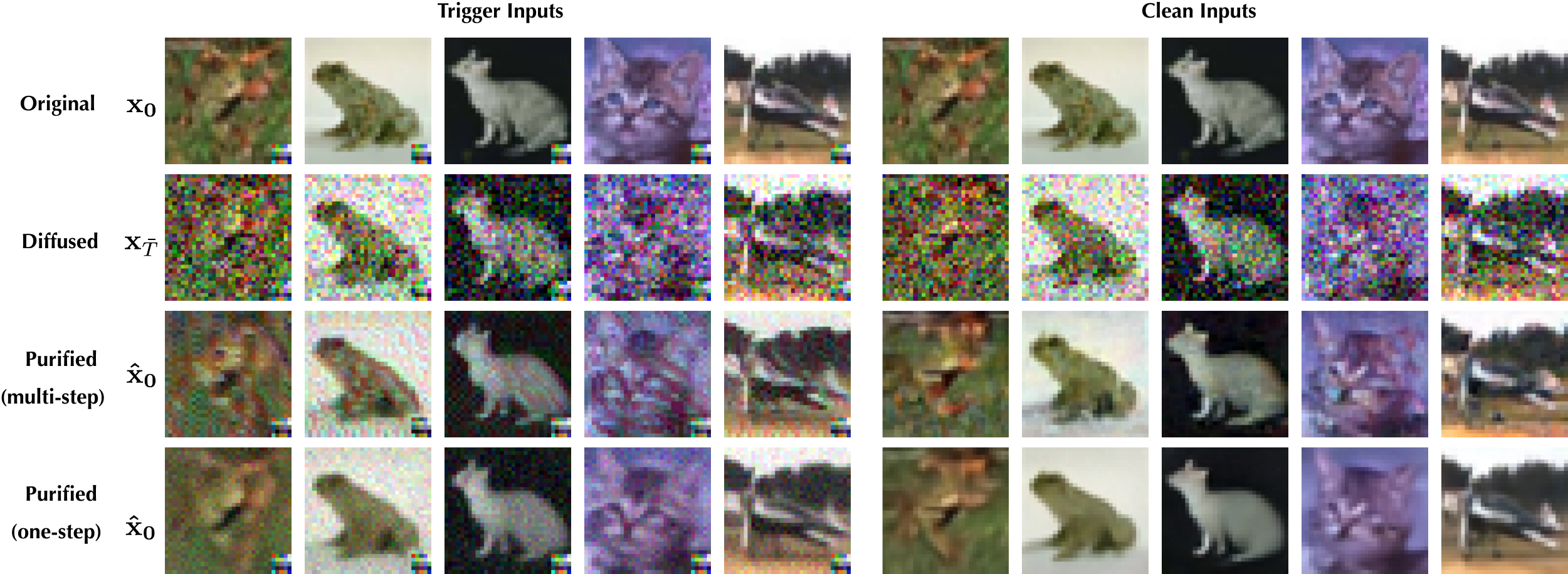}
    \caption{Original, diffused, and purified variants of clean and trigger inputs in \attack with non-adversarial triggers.}
    \label{fig:sample2}
\end{figure*}

\begin{table}[!ht]\footnotesize
\centering
\renewcommand{\arraystretch}{1.1}
\setlength{\tabcolsep}{3pt}
 {\footnotesize
\begin{tabular}{cc|cc|cc}
    &  & \multicolumn{2}{c|}{ASR (w/ $f \circ \phi$)} & \multicolumn{2}{c}{ASR (w/ $f \circ \phi^\sstar$)}                           \\ \cline{3-6} 
                           &                         &     Untargeted           &      Targeted            & Untargeted & Targeted  \\ \hline
                           \hline
\multirow{4}{*}{\rotatebox[origin=c]{90}{\parbox{1cm}{\centering Diffusion \\ Model}}}
& DDPM   & 11.6\%   &        10.4\%       &        87.3\%      &   73.8\%           \\
& DDIM   &  10.8\% &           9.8\%       &         80.2\%           &      44.3\%            \\
& SDE   &  7.9\% &    10.5\%           &        85.2\%      &  41.4\%     \\
& ODE  &  6.9\%\%   &   10.4\%        &        86.2\%        &     39.7\%          \\ \hline 
\multirow{3}{*}{\rotatebox[origin=c]{90}{Dataset}}
& CIFAR-10  & 11.6\% & 10.4\% &  87.3\% &  73.8\% \\
& CIFAR-100 & 41.6\% &  0.8\%&  95.8\%  &  70.9\%       \\
& CelebA    &  27.2\% &  18.7\% &  81.7\%  &  71.0\%
\end{tabular}}
\caption{\footnotesize Attack effectiveness of \attack with non-adversarial triggers ($f$: classifier only; $f \circ \phi$: classifier + benign diffusion model; $f \circ \phi^\star$: classifier + backdoored diffusion model). \label{table:effectiveness_asy}}
\end{table}

We evaluate the attack effectiveness and utility retention of \attack with non-adversarial triggers (defined as 5$\times$5 patch at the lower right corner), with results reported in Table\mref{table:effectiveness_asy} and Table\mref{table:utility_asy}. 
Figure\mref{fig:sample2} visualizes randomly sampled trigger and clean inputs, their latents, and their purified counterparts.
Observe in Table\mref{table:effectiveness_asy} that \attack with non-adversarial triggers is effective in both untargeted and targeted attacks. For example, against the DDPM model on CIFAR-10, \attack attains 73.8\% ASR in targeted attacks. However, \attack with non-adversarial triggers has a large impact on the diffusion model's utility as shown in Table\mref{table:utility_asy}. For example, across all the models, the clean ACC drops to around 10\%; in contrast, \attack with adversarial triggers has little influence on the diffusion model utility (cf. Table\mref{table:utility2}). Further, the non-adversarial trigger often causes the model's training to collapse. We speculate that this is because the forward and reverse processes of trigger inputs are `asymmetrical' (i.e., the trigger input $x_r$ as the input to the forward process and the adversarial input $\tilde{x}_r$ as the output of the reverse process), which tends to interfere with the normal diffusion process.




\begin{table}[!t]\footnotesize
\centering
\renewcommand{\arraystretch}{1.1}
\setlength{\tabcolsep}{2pt}
\begin{subtable}{\linewidth}
\centering
\begin{tabular}{c|c|cccc}
& \multirow{2}{*}{\attack} & \multicolumn{4}{c}{Diffusion Model} \\ 
\cline{3-6}
& & DDPM & DDIM & SDE & ODE \\
\hline
\multirow{2}{*}{Clean ACC} & w/o & 89.2\%  &  91.3\% & 91.8\% & 93.0\% \\   
& w/ & 80.6\% & 82.2\%  & 70.5\% & 81.2\% \\ 
\hline
Robust ACC  & w/o &  86.3\% & 82.1\%  & 86.5\%  &  79.6\% \\   
(PGD) & w/ & 75.3\% & 81.4\% &  60.3\% & 62.5\% \\ 
\hline
Robust ACC & w/o & 86.1\% &  82.5\% & 86.3\% & 78.3\% \\   
(AutoAttack) & w/ & 76.4\% & 80.9\% & 58.7\% &  64.5\% \\ 
\end{tabular}
\end{subtable}

\medskip 
\begin{subtable}{\linewidth}
\centering
\begin{tabular}{c|c|ccc}
& \multirow{2}{*}{\attack} & \multicolumn{3}{c}{Dataset} \\ 
\cline{3-5}
& & CIFAR10 & CIFAR100 & CelebA \\
\hline
\multirow{2}{*}{Clean ACC} & w/o & 89.2\%  & 61.1\%  & 75.4\% \\   
& w/ & 80.6\% & 57.8\% &  72.4\% \\ 
\hline
Robust ACC  & w/o & 86.3\%  & 51.2\%  & 42.7\% \\   
(PGD) & w/ & 75.3\% & 26.7\%  & 31.5\% \\ 
\hline
Robust ACC & w/o & 86.1\%  & 51.0\% & 40.5\% \\   
(AutoAttack) & w/ & 76.4\% & 25.8\%  & 32.6\%  \\ 
\end{tabular}
\end{subtable}
\caption{\footnotesize Utility retention of \attack with non-adversarial triggers (w/o: $f \circ \phi$ classifier + benign diffusion model; w/: $f \circ \phi^\star$ classifier + backdoored diffusion model). \label{table:utility_asy}}
\end{table}

Besides patch-based triggers\mcite{badnet,trojannn,baddiffusion,trojdiff}, we also evaluate other non-adversarial triggers, including blending-based\mcite{chen:2017:arxiv} and warping-based\mcite{wanet} triggers. Specifically, the blending-based attack generates a trigger input by blending a clean input with the trigger pattern (\meg, `Hello Kitty'), while the warping-based attack defines a specific image warping transformation (\meg, thin-plate splines) as the trigger and generates a trigger input by applying this transformation over a clean input. Based on the trigger designs in the original papers, we evaluate \attack on DDPM over CIFAR-10 under the default setting, with results summarized in Table\mref{table:alternative}.

\begin{table}[!ht]\footnotesize 
\centering
\renewcommand{\arraystretch}{1.1}
\begin{tabular}{c|cc|cc}
\multirow{2}{*}{Trigger} & \multicolumn{2}{c|}{Untargeted Attack} & \multicolumn{2}{c}{Targeted Attack} \\ \cline{2-5} 
                         & ACC            & ASR            & ACC           & ASR          \\ \hline
Blending-based                 & 81.6\%         & 47.2\%         & 77.6\%        & 20.1\%       \\
Warping-based                    & 81.3\%         & 64.3\%         & 78.4\%        & 36.7\%      
\end{tabular}
\caption{\footnotesize Evaluation of alternative trigger designs.\label{table:alternative}}
\end{table}

Notably, the alternative triggers are modestly effective under both targeted and untargeted settings. Meanwhile, they tend to produce less perceptible perturbations in purified inputs, as visualized in Figure\mref{fig:sample3} in \msec{sec:additional}. However, similar to patch-based triggers, they also result in lower attack effectiveness and large clean ACC drop. Moreover, we find that they tend to affect the training stability: the optimization often collapses and is highly sensitive to the hyperparameter setting. This may be explained by that these trigger patterns are more susceptible to being obscured by the diffusion process, leading to an entanglement between clean and trigger inputs in the latent space and, consequently, unstable training. 



\vspace{2pt}
{\bf Input-specific triggers --}
Recall that \attack uses a universal adversarial trigger across different inputs. We now explore the possibility of implementing input-specific triggers in \attack. Specifically, for each input $x$, we apply the PGD attack to generate its specific trigger $r$ as $x_r = (1-\alpha) x + \alpha r$. Then, similar to \attack, we train a backdoored diffusion model using these trigger inputs.

\begin{table}[!ht]\footnotesize 
\centering
\renewcommand{\arraystretch}{1.1}
\begin{tabular}{c|cc}
\multirow{2}{*}{Attack} & \multicolumn{2}{c}{Metric} \\ 
\cline{2-3} 
                         & Clean ACC            & ASR                    \\ \hline
Untargeted                &   47.6\%       &    59.2\%           \\
Targeted                    &    26.1\%     &     26.3\%      
\end{tabular}
\caption{\footnotesize \attack's attack performance with input-specific triggers.\label{table:dynamic}}
\end{table}

Table\mref{table:dynamic} evaluates \attack's performance with input-specific triggers against DDPM on CIFAR-10. Observe that although input-specific triggers also lead to effective attacks, they tend to considerably impact the clean accuracy. We also find that the training of diffusion models often collapses under such settings. We speculate that this is mainly due to the resemblance between input-specific triggers and random noise, which tend to interfere with the normal diffusion process of clean inputs. The manual inspection of post-purification clean inputs shows that these samples carry considerable random noise, which validates our specification.

\vspace{2pt}

\subsection{Existing attacks on diffusion models}

Given their focus on generative tasks and the necessity to activate the backdoor in the latent space, existing backdoor attacks\mcite{baddiffusion,trojdiff} on diffusion models cannot be directly applied to our setting. In particular, adapting BadDiffusion\mcite{baddiffusion} proves challenging as it is designed to link the trigger in the latent space to a specific output. However, it is possible to adapt TrojDiff\mcite{trojdiff} to our context. Specifically, we consider two possible schemes: i) Patching scheme, which diffuses a clean input $x$ and reverses it to an adversarial variant of its corresponding trigger input $x_r$. 
\begin{equation}
 x_t  = \sqrt{\bar{\alpha}_t}x + \sqrt{1 -\bar{\alpha}_t}( \gamma \epsilon + r)
 \label{trodiff_e}
\end{equation}
To implement this idea, we use \meq{trodiff_e} to substitute line 7 in Algorithm\mref{alg:attack} and keep the other setting the same as \attack. ii) Adversarial scheme, which samples the input from  $p^\sstar$ and reverses to itself. To this end, in addition to replacing line 7 in Algorithm\mref{alg:attack} with \meq{trodiff_e}, it is imperative to ensure that $\epsilon^\sstar = \epsilon$. 
\begin{table}[!t]{\footnotesize 
\centering
\renewcommand{\arraystretch}{1.1}
\begin{tabular}{c|cc}
Scheme      & Clean ACC & ASR \\ \hline
Patching       &   59.0\%        &  61.1\%   \\
Adversarial &    23.4\%       &    76.5\%
\end{tabular}
\caption{\footnotesize Evaluation of adapted TrojDiff.\label{table:trodiff_result}}}
\end{table}
Table\mref{table:trodiff_result} reports the evaluation results of these two schemes. Observe that TroDiff is much less effective than \attack. We speculate this is attributed to the generative formulation of TrojDiff, which only allows to activate the backdoor in the input space {\em approximately}.

\subsection{\attack-specific Defenses}



Here, we explore a \attack-specific defense that leverages the unique properties of diffusion models.

Before delving into details, we introduce the rationale behind this defense. Specifically, as a clean input $x$ and its trigger counterpart $x_r$ differ only by the trigger $r$, running the diffusion process on both $x_r$ and $x$ for a sufficiently large timestep $t$ results in the convergence of their respective latents, denoted by $\mathsf{diff}(x, t)$ and $\mathsf{diff}(x_r, t)$, which can be analytically proved: 
\begin{theorem}
Given $x_r \sim q_\mathrm{trigger}$ and its clean counterpart $x \sim p_\mathrm{data}$, let $q_t$ and $p_t$ be the distributions of $\mathsf{diff}(x_r, t)$ and $\mathsf{diff}(x, t)$, respectively. We have:
\begin{equation}
\frac{\partial D_\mathrm{KL}(p_t \| q_t)}{\partial t} \leq 0
\end{equation}
\end{theorem}
The proof (\msec{sec:proof}) follows\mcite{mle-sde,diffpure} while generalizing to both discrete and continuous diffusion models. 
Thus, the KL divergence between $q_t$ and $p_t$ consistently diminishes as $t$ increases throughout the diffusion process, suggesting that by increasing $t$, $\mathsf{diff}(x, t)$ becomes a close approximation of $\mathsf{diff}(x_r, t)$. Consequently, given a sufficiently large $t$, using $\mathsf{diff}(x, t)$ as the input for the reverse process, it is likely that the reverse process may yield samples from $q_\mathrm{trigger}$.

\begin{figure}[!t]
    \centering
    \includegraphics[width=0.4\textwidth]{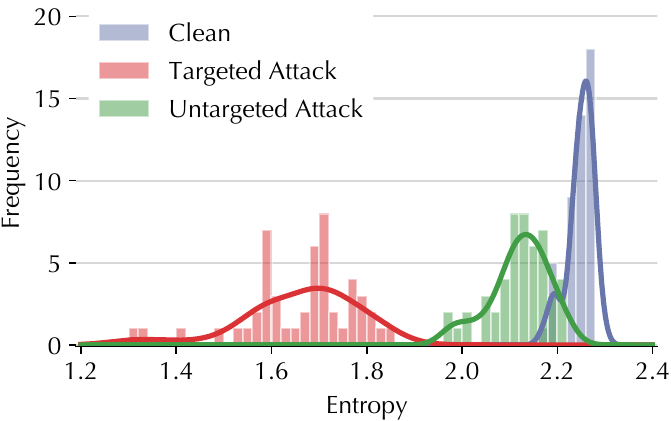}
    \caption{\footnotesize Entropy distributions of different diffusion models. }
    \label{fig:defense}
\end{figure}

Based on this insight, we propose the following defense. We randomly sample clean inputs from a reference dataset and feed them to the diffusion model. For each input $x$, we run the forward process for a sufficiently large timestep (e.g., 1,000) and the reverse process on the diffused input, which yields the output $\hat{x}$. We feed all $\{\hat{x}\}$ to the target classifier $f$ and measure the entropy of its predictions $\gH[\{f(\hat{x})\}]$. For a benign diffusion model, as $\hat{x}$ is likely to be mapped to $p_\mathrm{data}$, the entropy tends to be large; for a backdoored diffusion model, as $\hat{x}$ is likely to be mapped to $q_\mathrm{trigger}$, the entropy tends to be small, given that trigger inputs are designed to misclassified to the target class. To validate our hypothesis, we randomly sample 10 benign and 10 backdoored diffusion models. For each model, we evaluate the entropy of each model 5 times, with 100 inputs in each trial. Figure\mref{fig:defense} illustrates the entropy distributions of different models.

Observe that there exists a discernible difference in the entropy measures of benign and backdoored models, which is especially evident for backdoored models under targeted attacks.  
While the distribution of untargeted attacks overlaps with that of benign models, they remain distinguishable. This finding validates our hypothesis, highlighting entropy as a critical discriminative measure for detecting backdoored diffusion models. Further, by examining the classification outcomes and reconstructed outputs, it is possible to reverse-engineer the potential trigger patterns.


While appearing promising, this defense also suffers significant limitations. First, a notable entropy difference between clean and backdoored diffusion models is evident only when the reconstructed input $\hat{x}$ is adversarial (i.e., the trigger distribution $q_{trigger}$ is adversarial). Thus, non-adversarial triggers, such as those discussed in \msec{negative}, will not cause a notable entropy difference. Further, reverse-engineering the trigger becomes challenging when the trigger pattern is invisible, as exemplified by the warping-based trigger\msec{negative}, due to its invisibility and non-adversarial nature. Finally, note that the entropy distributions of clean and (untargeted) backdoored models overlap significantly, indicating the challenge of defending against untargeted \attack. We consider developing a full-fledged defense against \attack as our ongoing research.

%% file: literature.tex
\section{Related Work}


{\bf Diffusion models --} The recent advances in diffusion models\mcite{ddpm,sde,consistency-model,guided-diffusion} have led to breakthroughs across a variety of generative tasks such as image generation\mcite{old-sde,ddpm,ddim}, audio synthesis\mcite{audio-synthesis}, and density estimation\mcite{density-estimation}. More recently, due to their remarkable denoising capabilities, diffusion models have been utilized to defend against adversarial attacks\mcite{szegedy:iclr:2014,goodfellow:fsgm} via purifying adversarial inputs\mcite{diffpure, sde-diffpure} or improving certified robustness\mcite{carlini2022certified,densepure}. 
However, there is still a lack of understanding about the vulnerability of diffusion models, which is concerning given the increasing use of pre-trained diffusion models in security-enhancing use cases.

{\bf Backdoor attacks and defenses --} 
As a major threat to machine learning security, backdoor attacks implant malicious functions into a target model during training, which are activated via trigger inputs at inference. Many backdoor attacks have been proposed in the context of classification tasks, which can be categorized along i) attack targets -- input-specific\mcite{poisonfrog}, class-specific\mcite{tact}, or any-input\mcite{badnet}, ii) trigger visibility -- visible\mcite{badnet, chen:2017:arxiv, saha2022backdoor} and imperceptible\mcite{wanet, ctrl} triggers, and iii) optimization metrics -- attack effectiveness\mcite{imc}, transferability\mcite{latent-backdoor,xi2023security}, model architecture \cite{pang2022dark} or  evasiveness\mcite{poisonfrog}. Meanwhile, in generative tasks, the adversary aims to generate outputs from a specific distribution\mcite{trojanlm,gan-backdoor}. To mitigate such threats, many defenses have also been proposed, which can be categorized according to their strategies: i) input filtering purges poisoning inputs from the training data\mcite{tran:2018:nips,chen2018detecting}; 
ii) model inspection determines whether a given model is backdoored\mcite{kolouri2020universal, huang2019neuroninspect, abs, wang2019neural}; iii) input inspection detects trigger inputs at inference time\mcite{tact, gao2019strip}; and iv) model sanitization modifies (e.g., pruning) the model to remove the  backdoor\mcite{anp,zheng2022data}. However, it is found that given defenses are often circumvented or even penetrated by stronger or adaptive attacks\mcite{backdoor-revisiting,xi2023defending}, leading to a constant arms race between attackers and defenders.

{\bf Backdoor attacks on diffusion models --} 
The prohibitive training costs of diffusion models often force users to rely on pre-trained, ready-to-use models, making them vulnerable to backdoor attacks. TrojDiff\mcite{trojdiff} and BadDiffusion\mcite{baddiffusion} explore backdoor attacks in this context, which focus on the reverse process of the diffusion models. Specifically, these attacks force the diffusion model to generate specific outputs by attaching the trigger to the sampled Gaussian noise to activate the backdoor in the latent space, which is generally infeasible in many real-world applications (e.g., adversarial purification) since the adversary has no control over the reverse process. Additionally, these attacks only explore the security vulnerability of diffusion models as standalone models.

To the best of our knowledge, this work is the first one on backdoor attacks tailored to security-centric diffusion models, aiming to diminish the security assurance of diffusion models to the foundation models by activating backdoors in the input space. 

%% file: conclusion.tex
\section{Conclusion}
\label{sec:conclusion}

This work studies the potential risks of using pre-trained diffusion models as defensive tools in security-enhancing use cases. We present \attack, a novel backdoor attack that integrates malicious forward-reverse processes into diffusion models, guiding trigger inputs toward adversary-defined distributions. By exploiting such diffusion backdoors, the adversary is able to significantly diminish the security assurance provided by diffusion models in use cases such as adversarial purification and robustness certification. Our findings raise concerns about the current practice of diffusion models in security-enhancing contexts and shed light on developing effective countermeasures.

%% file: appendix.tex

\subsection{Default parameter setting}
\label{sec:param}
Following prior work\mcite{trojdiff}, to reduce the training cost, we use pre-trained diffusion models and apply \attack to fine-tune them.  Table\mref{table:default} summarized the default parameter setting of \attack.

\begin{table}[!ht]{\footnotesize
\centering
\renewcommand{\arraystretch}{1.1}
\setlength{\tabcolsep}{2pt}

\centering
\begin{tabular}{c|c|c}
Type                                  & Parameter & Setting                                                               \\ \hline
\multirow{7}{*}{Trigger generation} & Trigger  size                           &   \multicolumn{1}{c}{\begin{tabular}[c]{@{}c@{}}$32\times 32$ for CIFAR-10/-100\\  $64\times 64$ for  CelebA (64$\times$64)\end{tabular}}                                                                                   \\
                                                                                                                                
                                        & Surrogate classifier                &    ResNet-18                                                                                                                  \\
                                                                                                                            
                                        & Mixing weight     $\alpha$                  &       0.05 \\
                                         & Optimizer                  &       Adam  \\
                                         & Learnining rate                 &    \(1 \times 10^{-1}  \) \\
                                         & Target class                &   0 if applicable \\ \hline
\multirow{7}{*}{Fine-tuning}            & Optimizer                           & \multicolumn{1}{c}{Adam}                                                                                            \\
                                        & Learning rate                       & \multicolumn{1}{c}{\(2 \times 10^{-4}\)}                                                                            \\
                                        & Epochs                              & \multicolumn{1}{c}{\begin{tabular}[c]{@{}c@{}}20 for CIFAR-10/-100\\  10 for CelebA (64$\times$64)\end{tabular}} \\
                                        & Batch size                          & \multicolumn{1}{c}{\begin{tabular}[c]{@{}c@{}}256 for CIFAR-10/-100\\ 64 for CelebA (64 $\times$ 64)\end{tabular}} \\
                                        & Truncated timestep                       & \multicolumn{1}{c}{300}                                                                                         

\end{tabular}
\caption{\footnotesize Default parameters of \attack. \label{table:default}}}
\end{table}









\subsection{Proof}
\label{sec:proof}

\subsubsection{Proof of Theorem 3.1}
To simplify the analysis, we define the trigger $r$ as a shift applied on clean inputs (i.e.,  
$x_r = x + r$). Let $q(x)$ be the distribution of clean data $\gD$, on which the benign diffusion model is trained. Let $q_r$ denote the distribution $q(x - r)$ for $x \sim \gD$. Also, let $\hat{p}$ be the output distribution when the input of the backward process is a linear combination $(1-\alpha)x_r + \alpha \epsilon$. We thus have the following derivation.
\begin{flalign*}
    &D_\mathrm{KL}\left(q(x-r) \middle\| \hat{p}(x)\right) - D_\mathrm{KL}\left(q(x) \middle\| \hat{p}(x)\right) &\\
    &= \int q(x-r) \log \frac{q(x-r)}{\hat{p}(x)}  \dd x - \int q(x) \log \frac{q(x)}{\hat{p}(x)}  \dd x &\\
    &= \int q(x) \log \frac{q(x)}{\hat{p}(x+r)}  \dd x - \int q(x) \log \frac{q(x)}{\hat{p}(x)}  \dd x & \\
    &= \int q(x) \log \frac{\hat{p}(x)}{\hat{p}(x+r)}  \dd x & \\
    &= -\int q(x) \log \frac{\hat{p}(x+r)}{\hat{p}(x)}  \dd x & \\
    &= -\int q(x) \log \frac{\hat{p}(x)+\nabla\hat{p}(x)\cdot r + o(\|r\|^2)}{\hat{p}(x)}  \dd x & \\
    &= -\int q(x) \left( \nabla\log\hat{p}(x) \cdot r \right) \dd x + o(\|r\|^2) & \\
    &= -\mathbb{E}\left[  \nabla\log\hat{p}(x) \cdot r \right] + o(\|r\|^2) &
\end{flalign*}
According to Theorem 1 in\mcite{ood-shift}, we have
\begin{flalign*}
    & D_\mathrm{KL}\left(q_r \middle\| \hat{p}\right) = D_\mathrm{KL}\left(q \middle\| \hat{p}\right) -\mathbb{E}\left[  \nabla\log\hat{p} \cdot r \right] + o(\|r\|^2) & \\
    &\leq \mathcal{J}_{\mathrm{SM}} + D_\mathrm{KL}\left(q_T \middle\| \rho\right) + \mathcal{F}(\alpha) -\mathbb{E}\left[  \nabla\log\hat{q} \cdot r \right] + o(\|r\|^2) &
\end{flalign*}
where $\mathcal{J}_{\mathrm{SM}}$ is the weighted score matching loss, $q_T$ is the distribution at time $t$ in the forward transformation, $\rho$ is the distribution of standard Gaussian noise, $\mathcal{F}(\alpha)$ is introduced by the distribution of the OOD testing samples, which is controlled by the forward noise weight $\alpha$ and converges to 0 as
$\alpha$ goes to 1.

\subsubsection{Forward}

To facilitate implementing \attack for various diffusion models, we unify and simplify the notations of discrete (\meg, DDPM~\mcite{ddpm}) and continuous (\meg, SDE~\mcite{sde}) diffusion models.

\textbf{Discrete}
For $\forall t \in \mathbb{Z}^+$, we have the one-step relation:
\begin{equation*} 
x_t = \sqrt{\rvalpha_t} x_{t-1} + \sqrt{1-\rvalpha_t} \rvepsilon_{t-1,t}
\end{equation*}
where $0<\rvalpha_t<1$. Extend it to multi-step ($t' \geq t$):
\begin{equation*} 
\label{eq:multi-step}
x_{t'} = \sqrt{\prod^{t'}_{\tau=t} \rvalpha_\tau} x_t + \sqrt{1-\prod^{t'}_{\tau=t} \rvalpha_\tau} \rvepsilon_{t, t'}
\end{equation*}
We define the product of $\rvalpha_t$ as $\bar{\rvalpha}_t$:
\begin{equation*} 
\bar{\rvalpha}_t = \begin{cases}
\prod^{t}_{\tau=1} \rvalpha_\tau,\quad &t>0 \\
1,\quad &t=0
\end{cases}
\end{equation*}
Based on $0<\rvalpha_t<1$, we have $0<\bar{\rvalpha}_T<\bar{\rvalpha}_{T-1}<\cdots<\bar{\rvalpha}_0=1$.
Therefore, the previous \meq{eq:multi-step} could be reformalized as
\begin{equation*} 
x_{t'} = \sqrt{\frac{\bar{\rvalpha}_{t'}}{\bar{\rvalpha}_t}} x_t + \sqrt{1-\frac{\bar{\rvalpha}_{t'}}{\bar{\rvalpha}_t}} \rvepsilon_{t, t'}
\end{equation*}
A more symmetric form is
\begin{equation*} 
\label{eq:forward}
\frac{x_{t'}}{\sqrt{\bar{\rvalpha}_{t'}}} - \frac{x_t}{\sqrt{\bar{\rvalpha}_t}} = \sqrt{\frac{1}{\bar{\rvalpha}_{t'}} - \frac{1}{\bar{\rvalpha}_t}}\rvepsilon_{t, t'}, \quad t'\geq t
\end{equation*}
Replace with new variables:
\begin{equation} 
\label{eq:transform}
\begin{cases}
s_t = \frac{1}{\bar{\rvalpha}_{t}} - \frac{1}{\sqrt{\bar{\rvalpha}_0}}, \quad &t\in \mathbb{Z}^+ \\
y_{s_t} = \frac{x_t}{\sqrt{\bar{\rvalpha}_t}} - \frac{x_0}{\sqrt{\bar{\rvalpha}_0}} \\
\end{cases}
\end{equation}
We have $s_T>s_{T-1}>\cdots>s_0 = 0$, and
\begin{equation*} 
\label{eq:forward_s}
\begin{cases}
y_0 = 0 \\
y_{s'} - y_s = \sqrt{ s' - s }\rvepsilon_{s, s'} \sim \mathcal{N}(0, s'-s) , \quad &s'\geq s\\
\end{cases}
\end{equation*}

\textbf{Continuous}
When $t$ is extended to $[0,+\infty)$ and $\bar{\rvalpha}_t$ is assumed to be a monotonically decreasing continuous function where $\lim\limits_{t\to\infty} \bar{\rvalpha}_t =0$, we could extend $s$ to $[0,+\infty)$ as well. From \meq{eq:forward_s}, we know $y_s$ is a Wiener process.

\subsubsection{Proof of Theorem 5.1}

{\scriptsize
\begin{align*}
    \frac{\partial D_\mathrm{KL}(p_t \| q_t)}{\partial s} &= \frac{\partial}{\partial s} \int{p(y_s) \log \frac{p(y_s)}{q(y_s)} \dd y} \\
    &= \int{\left( \frac{\partial p(y_s)}{\partial s}  \log \frac{p(y_s)}{q(y_s)} + \frac{\partial p(y_s)}{\partial s} + \frac{\partial q(y_s)}{\partial s} \frac{p(y_s)}{q(y_s)}\right)}\dd y
\end{align*}}
The integration of second term is 0. Wiener process $y_s$ satisfies
{\scriptsize
\begin{equation*}
\frac{\partial p(y_s)}{\partial s} = \frac{1}{2} \frac{\partial^2 p(y_s)}{\partial y_s^2}
\end{equation*}}
We have
{\scriptsize
\begin{equation*}
    \frac{\partial D_\mathrm{KL}(p_s \| q_s)}{\partial s} = \frac{1}{2}\int{\left( \frac{\partial^2 p(y_s)}{\partial y_s^2}  \log \frac{p(y_s)}{q(y_s)} + \frac{\partial^2 q(y_s)}{\partial y_s^2} \frac{p(y_s)}{q(y_s)}\right)}\dd y
\end{equation*}}
Using integration by parts, we have the following derivation:
{\scriptsize
\begin{align*}
    \frac{\partial D_\mathrm{KL}(p_s \| q_s)}{\partial s} = & -\frac{1}{2}\int{\left( \frac{\partial p(y_s)}{\partial y_s}  \frac{\partial \log \frac{p(y_s)}{q(y_s)}}{\partial y_s} + \frac{\partial q(y_s)}{\partial y_s} \frac{\partial \frac{p(y_s)}{q(y_s)}}{\partial y_s}\right)}\dd y \\
     = & -\frac{1}{2}\int\left(p(y_s) \frac{\partial \log p(y_s)}{\partial y_s} \frac{\partial \log \frac{p(y_s)}{q(y_s)}}{\partial y_s} \right.\\
     & + \left.q(y_s) \frac{\partial \log q(y_s)}{\partial y_s} \frac{p(y_s)}{q(y_s)} \frac{\partial \log \frac{p(y_s)}{q(y_s)}}{\partial y_s}\right) \dd y \\
     = & -\frac{1}{2}\int{ p(y_s) \left( \frac{\partial \log \frac{p(y_s)}{q(y_s)}}{\partial y_s} \right)^2}\dd y \\
     = & -\frac{1}{2} \mathbb{E} \left[ \left( \frac{\partial \log \frac{p(y_s)}{q(y_s)}}{\partial y_s} \right)^2 \right]
\end{align*}}
Therefore, it is essentially the Fisher information
{\scriptsize
\begin{equation*}
    \frac{\partial D_\mathrm{KL}(p_t \| q_t)}{\partial s} = -\frac{1}{2} D_\mathrm{F}(p_t \| q_t) \leq 0
\end{equation*}}
According to the transformation between $s$ and $t$ in \meq{eq:transform} and the monotonicity of $\bar{\rvalpha}_t$,
{\scriptsize
\begin{equation*}
    \frac{\partial D_\mathrm{KL}(p_t \| q_t)}{\partial t} = \frac{\dd s}{\dd t} \frac{\partial D_\mathrm{KL}(p_t \| q_t)}{\partial s} = \frac{1}{2 \bar{\rvalpha}_t^2} \frac{\dd \bar{\rvalpha}_t}{\dd t} D_\mathrm{F}(p_t \| q_t) \leq 0
\end{equation*}}

\subsection{Additional Results}
\label{sec:additional}

\begin{figure*}[!ht]
    \centering
    \includegraphics[width=0.9\textwidth]{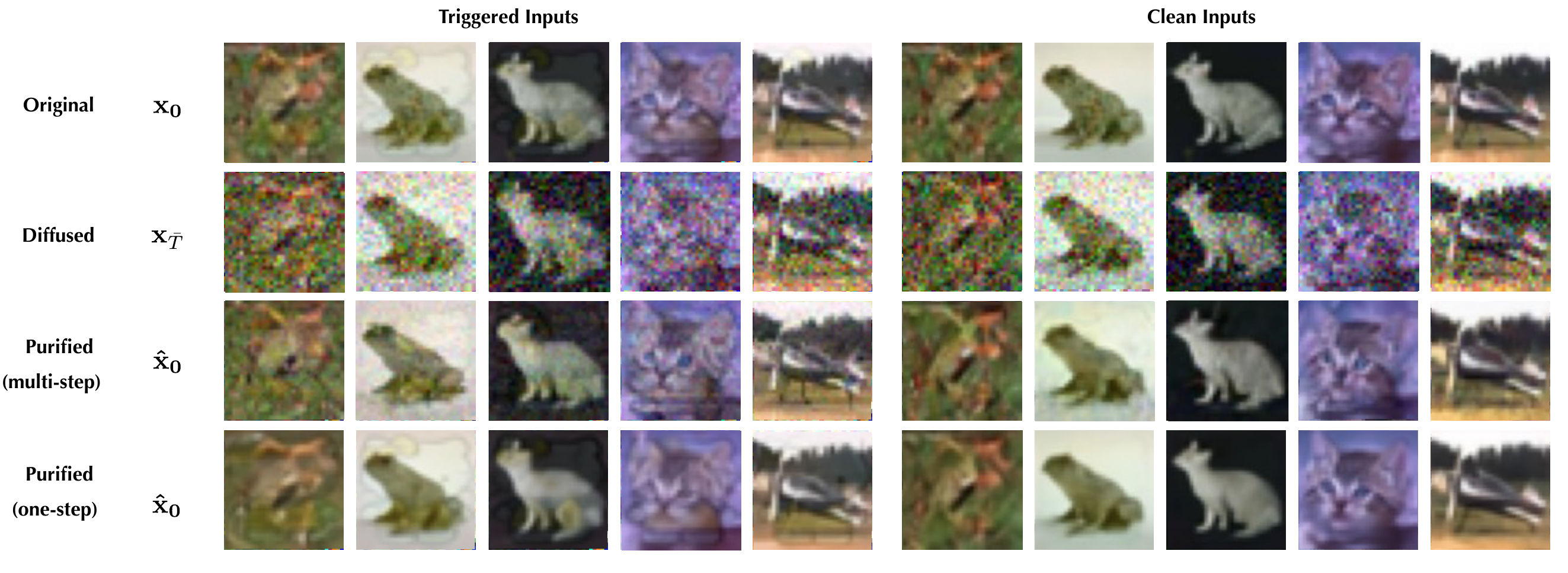}
    \includegraphics[width=0.9\textwidth]{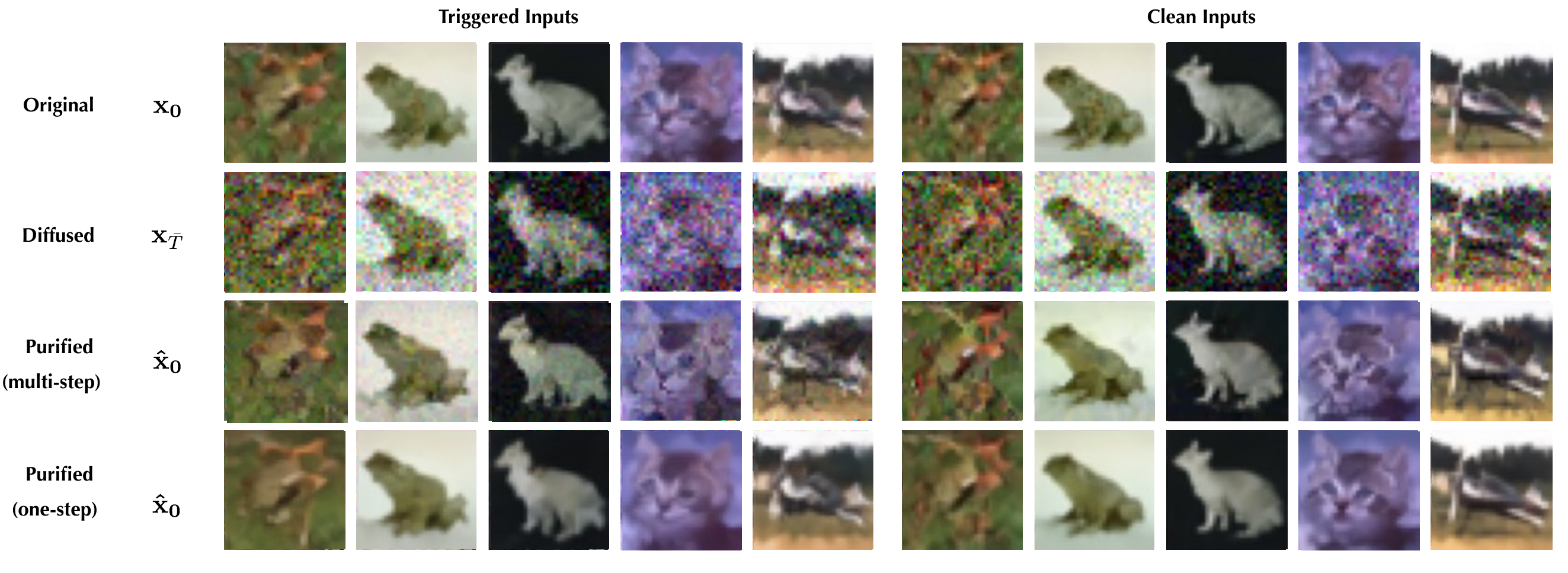}
    \caption{Original, diffused, and purified variants of clean and trigger inputs in \attack with blending triggers (upper) and warping triggers (lower).}
    \label{fig:sample3}
\end{figure*}




\subsubsection{Adversarial neuron pruning} 
\begin{figure}[!ht]
    \centering
    \includegraphics[width=0.5\textwidth]{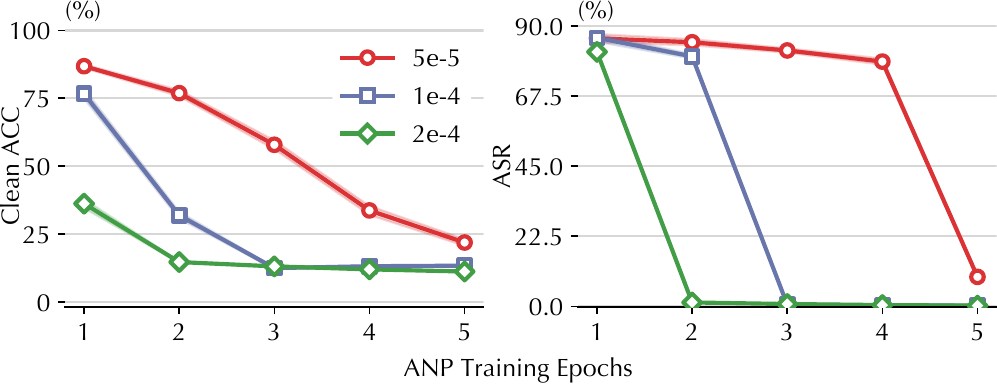}
    \caption{\footnotesize Effectiveness of ANP against \attack (targeted attacks) under varying learning rate.}
    \label{fig:anp}
\end{figure}

We further consider adversarial neuron pruning (ANP)\mcite{anp}, a pruning-based defense against backdoor attacks. Based on the assumption that neurons sensitive to adversarial perturbation are strongly related to the backdoor, ANP removes the injected backdoor by identifying and pruning such neurons. 
In\mcite{baddiffusion}, ANP is extended to the setting of diffusion models.
Following\mcite{baddiffusion}, we apply ANP to defend against (targeted) \attack on DDPM. We assume ANP has access to the full (clean) dataset and measure \attack's performance under varying ANP learning rates, with results summarized in Figure\mref{fig:anp}. We have the following interesting observations.


Overall, ANP is effective against \attack but at the substantial cost of clean accuracy. For example, when ANP's learning rate is set at 5e-5, a decrease in ASR below 20\% correlates with a significant drop in clean ACC, falling below 25\%. This trade-off is even more evident for larger learning rates. For instance, at a learning rate of 2e-4, while ASR approaches nearly zero, clean ACC also plummets to around 10\%. This can be explained as follows. In \attack's optimization (cf. \meq{eq:opt}), the backdoor diffusion process is deeply intertwined with the benign diffusion process. Recall that ANP attempts to eliminate the backdoor by separating neurons sensitive to the backdoor function. However, due to the entanglement between the backdoor and normal diffusion processes, pruning invariably affects the utility adversely.